\documentclass[aps,pre,reprint,superscriptaddress,showpacs]{revtex4-1}%
\usepackage{amsfonts}
\usepackage{amsmath}
\usepackage{amssymb}
\usepackage{graphicx}
\usepackage{color}
\usepackage{amsmath}

\begin{document}
\title{Controlling the energy of defects and interfaces in the amplitude expansion of the phase-field crystal model} \author{Marco Salvalaglio}
\email{marco.salvalaglio@tu-dresden.de} 
\affiliation{Institute  of Scientific Computing,  Technische  Universit\"at
Dresden,  01062  Dresden,  Germany} \author{Rainer Backofen} 
\affiliation{Institute  of Scientific Computing,  Technische  Universit\"at
Dresden,  01062  Dresden,  Germany} \author{Axel Voigt}
\affiliation{Institute  of Scientific Computing,  Technische  Universit\"at
Dresden,  01062  Dresden,  Germany} \affiliation{Dresden Center for
Computational Materials Science (DCMS), TU Dresden, 01062 Dresden, Germany}
\author{Ken R. Elder} \affiliation{Department of Physics, Oakland University,
Rochester, 48309 Michigan, USA.}


\begin{abstract}
One of the major difficulties in employing phase field crystal (PFC) modeling and the associated amplitude (APFC) formulation is the ability to tune model parameters to match experimental quantities. In this work we address the problem of tuning the defect core and interface energies in the APFC formulation.  We show that the addition of a single term to the free energy functional can be used to increase the solid-liquid interface and defect energies in a well-controlled fashion, without any major change to other features.  The influence of the newly added term is explored in two-dimensional triangular and honeycomb structures as well as bcc and fcc lattices in three dimensions. In addition, a finite element method (FEM) is developed for the model that incorporates a mesh refinement scheme. The combination of the FEM and mesh refinement to simulate amplitude expansion with a new energy term provides a method of controlling microscopic features such as defect and interface energies while simultaneously delivering a coarse-grained examination of the system.
\end{abstract}

\maketitle

\section{Introduction}
\label{sec:introduction}
In the past few decades, phase-field (PF) models have been used extensively for modeling the ordering of nano- and micro-structures.  Such models provide a suitable framework for the investigation of a wide range of phenomena such as solidification processes, grain growth, surface diffusion, heteroepitaxy, and even dislocation dynamics \cite{Chen2002,Boettinger2002,Steinbach2009,Li2009,Bergamaschini2016b}.
Despite their versatility, strong limitations arise for PF models when looking at material properties closely related to atomic arrangement and periodicity. To account for these microscopic properties, the so-called phase-field crystal (PFC) model was developed \cite{Elder2002,Elder2004}. It consists of a continuum field theory that describes the local atomic probability density. Moreover, it allows one to cope with the dynamics of atomic structures at diffusive time scales so that the fast dynamics of vibration of atoms is filtered out \cite{Emmerich2012}.  The downside of the PFC approach is that the spatial resolution required in numerical simulations is determined by the lattice constant. Therefore, simulations of PFC models are restricted to systems much smaller than can be accessed in standard PF models.

To overcome the length scale limitation of PFC models, the amplitude expansion, also referred to as renormalization-group reduction, of the PFC model (APFC) \cite{Goldenfeld2005,Athreya2006,GoldenfeldJSP2006} was developed. It is based on the idea that the continuous density in PFC models can be described by the amplitude of the minimum set of Fourier modes or wave vectors needed for a given crystal symmetry.  To allow for crystals in arbitrary orientations, strained systems and/or defects, the amplitudes are complex functions.  Roughly speaking the magnitude of the amplitudes accounts for the liquid and solid phases, while the phase incorporates elasticity and crystals rotations.  The combination of the magnitude and phase allows for defects.  In this approach a coarser spatial resolution than standard PFC can be used, thus allowing for the simulation of much larger systems. Moreover, this representation enables the use of an optimized spatial discretization \cite{AthreyaPRE2007}.  

Simulations of the APFC model have been shown to be very useful for studying a wide variety of phenomena. The method has been applied to the study of polycrystalline films and the motion of grain boundaries (GBs) \cite{Goldenfeld2005,Athreya2006,GoldenfeldJSP2006,Yeon2010}, the study of heteroepitaxial ordering of ultrathin films \cite{Elder12,Elder13,Elder16b,Smirman17}, structural phase transitions \cite{Ofori13} and grain boundary energies in graphene \cite{Hirvonen2016}. The method has also been extended to binary systems \cite{ElderPRE2010,Huang10,Spatschek2010}. Moreover, it has been used to examine the influence of compositional strains on interfaces \cite{Huang16}, heteroepitaxy in binary systems \cite{ElderPRE2010,Huang10} and the elastically induced interaction of GBs and compositional interfaces \cite{Geslin2015,Xu2016}. While the original APFC model was introduced for two-dimensional systems with triangular symmetry the method has been extended to fcc and bcc systems in three dimensions \cite{ElderPRE2010,ElderJPCM2010} and honeycomb lattices in two dimensions \cite{Elder16b,Hirvonen2016}.  Other advances include exploiting the phase of amplitudes to achieve instantaneous mechanical equilibrium even under extreme conditions \cite{Heinonen2014}. Most of these investigations were performed with simulations using simple numerical methods on a fixed grid. In this paper we adopt a more advanced computational method, i.e., an adaptive finite element methods (FEM) with a semi-implicit integration scheme. 

The main purpose of this work is to propose a method to control the energies of dislocation cores and ordered-disordered interfaces in APFC models.  PFC and APFC models are similar to traditional PF models. They are both essentially long wavelength theories, i.e., only the lowest order gradients, or Fourier modes, are retained in the free energies that enter such models.  This implies that the predictions of such models on small length scales are not accurate. For example, the exact shape of domain walls (often described by hyperbolic tanh profiles in $\phi^4$ models \cite{Cahn1958}) in PF models or the density profiles near dislocation cores in PFC models are unlikely to match experimental systems. 

The validity of PF models, however, can be shown by taking the limit for vanishing thicknesses of the interfaces between phases and showing that they reduce to traditional sharp interface (SI) models \cite{Li2009}.  This matching is advantageous as it connects the parameters that enter continuous models with those that enter the SI models, which are typically well characterized in terms of known constants, such as surface tension, capillary lengths, diffusion constants etc.. A very important point is that although the predictions of the PF models on small length scales (i.e., interfacial or domain wall thicknesses) are qualitative, they can be used to make quantitative predictions on long length scales. The reason for this dichotomy is that the dynamics are strongly influenced by the existence of small length scale features, such as surfaces and dislocation cores, but not necessarily the exact spatial variation on small scales.  

In much the same way, PFC modeling can be thought of as a long-wavelength model, even though it creates structure on the atomic scales, as explicitly considered in the derivation via dynamical density-functional theory \cite{Elder2007,VanTeeffelen2009}. It is also straightforward to show that for small deformations, long wavelength limit PFC models reduce to continuum elasticity theory \cite{Elder2002,Heinonen2014}. Similarly, in binary PFC models, it is easy to show that they reduce to traditional phase field models of binary alloy solidification with elastic interactions, such as Vegard's law \cite{Huang10,ElderPRE2010}.  In addition, PFC models go beyond linear elasticity theory since they incorporate dislocations in a natural manner and can be shown to reproduce well-known results, such as the Read-Shockley equation for low angle GBs that consist of an array of dislocation cores \cite{Elder2002,Elder2004,Jaatinen2009,Hirvonen2016}.  While these results in some sense validate the PFC approach, it is difficult to match the original model to experimental systems.

The main reason for this difficulty is that the original PFC model essentially contains only two adjustable parameters, as obtained by rewriting the free energy in dimensionless units \cite{Elder2004} (i.e., by scaling to a dimensionless length, density, and temperature).  These parameters are related to temperature and the average density.  Clearly in a system that has, for example, several distinct elastic moduli, only one of them can be fitted exactly.  For example, in a three-dimensional (3D) bcc system, the original PFC model gives, $C_{11}=C_{22}=C_{33}$ and $C_{12}=C_{13}=C_{23}=C_{44}=C_{55}=C_{66}=C_{11}/2$, thus it is not possible to fit for example, $C_{11}$ and $C_{12}$ independently. This is a serious deficiency, although considering the lack of parameters that enter the original mode, it is not a surprising result.  Fortunately, adding more modes, or including higher order gradients, does lead to more flexibility in selecting the elastic moduli \cite{Hirvonen2016,greenwood11,mkhonta13,mkhonta16}.

Perhaps a more difficult problem in PFC modeling is controlling the defect core energies, which naturally will play a very important role in polycrystalline materials. The goal is not to accurately describe the structure of the cores (similarly to traditional PF modeling not accurately describing interfacial profiles in most cases), but to tune the cores to match experiments or other theoretical predictions.  

In this paper we consider adding a modification to APFC models such that the energy of solid-liquid interfaces and dislocation cores can be tuned.  An additional term in the free energy is considered, which is non-vanishing when the order of the solid phases changes. A similar approach has been recently proposed for the PFC model in order to include phase transition \cite{Kocher2015} and to introduce an adjustable interface energy \cite{Guo2016}. Here we propose a suitable formulation to account for these
effects in APFC models, exploiting an order parameter directly connected to the amplitude functions. 

The work is organized as follows. In Sec.~\ref{sec:APFC} the standard APFC approach is reported, highlighting its generality with respect to the symmetry of the crystalline phase. Then, the additional term in the free energy allowing for a tuning of the energy of defects and interfaces is introduced in Sec.~\ref{sec:coreenergy}. In Sec.~\ref{sec:num} the main features of the numerical method adopted in this work are illustrated. The effect of the newly-introduced energy term on the shape and the energetics of solid-liquid interfaces is discussed in Sec.~\ref{sec:surf}. The results concerning tuning the core-energy
of defects forming at straight GBs between tilted crystals are addressed in Sec.~\ref{sec:tilted}, focusing on the case of 2D honeycomb structures. The possibility to control the energy of GBs as a whole is also illustrated therein. Section \ref{sec:strained} addresses the control of the energy of defects in multilayered strained systems, where both 2D and 3D symmetries are explicitly considered. Conclusions and remarks are given in Sec.~\ref{sec:conclusions}. The symmetry-dependent terms in the APFC equations, the time-integration scheme and additional details concerning some specific setups for simulations are reported in the Appendixes.

\section{Model}
\label{sec:APFC}

	The free energy functional, $F_n$, in the PFC model can be written in terms of the dimensionless density difference, $n$, in the following form:
\begin{equation}
F_n=\int_{\Omega} \left[\frac{\Delta B_0}{2}n^2+\frac{B^x_0}{2} n(1+\nabla^2)^2n 
-\frac{t}{3}n^3+\frac{v}{4}n^4 \right]d\mathbf{r},
\end{equation}
where, $\Delta B_0$, $B_0^x$, $v$, and $t$ are parameters that control the phase diagram and properties of the system; see \cite{Elder2007}.  This free energy describes a first order phase transition from a disordered or liquid state ($n$ constant) at high $\Delta B_0$ to a crystalline state ($n$ periodic) at low or negative $\Delta B_0$. In \cite{Goldenfeld2005,Athreya2006,GoldenfeldJSP2006} it is shown that a so-called amplitude expansion can be derived by coarse-graining the density $n$.  In this approach, $n$ is written as
\begin{equation}
n=n_0+\sum_{j=1}^N \left[ \eta_j(\mathbf{x},t) e^{i\mathbf{k}_j \cdot
\mathbf{x}}+ \eta_j^*(\mathbf{x},t) e^{-i\mathbf{k}_j \cdot \mathbf{x}}\right],
\label{eq:density}
\end{equation}
where $N$ is the number of reciprocal-lattice vectors $\mathbf{k}_j$ required to reproduce a specific symmetry ($N=3$ for 2D triangular or honeycomb symmetry, $N=6$ for bcc lattices, and $N=7$ for fcc lattices; see Ref.~\cite{ElderPRE2010}). The $\mathbf{k}_j$ vectors for the lattices considered in
this work are reported in Appendix \ref{app:A}.

The $\eta_j$'s are the complex amplitude functions.  With the exception of \cite{Yeon2010} and \cite{Huang10}, the average $n$ (i.e., $n_0$) is assumed to be constant in space, and with an appropriate 
definition it can be set to zero without loss of generality \cite{ElderPRE2010}. Assuming that $\eta_j$ varies on length scales larger than the atomic spacing (i.e., $2\pi/|\mathbf{k}_j|$), the free-energy functional reads
\begin{equation}
\begin{split}
F=\int_{\Omega} &\bigg[\frac{\Delta B_0}{2}A^2+\frac{3v}{4}A^4 +\sum_{j=1}^N
\left ( B_0^x |\mathcal{G}_j \eta_j|^2-\frac{3v}{2}|\eta_j|^4 \right )
\\ & +f^s(\{\eta_j\},\{\eta^*_j\}) \bigg]  d \mathbf{r}, \end{split}
\label{eq:energyamplitude}
\end{equation}
where $\mathcal{G}_j\equiv \nabla^2+2i\mathbf{k}_j \cdot \nabla$ and $A^2\equiv 2\sum_{j=1}^N |\eta_j|^2$. $f^s(\{\eta_j\},\{\eta_j^*\})$ is set in agreement with the appropriate symmetry as reported in Appendix \ref{app:A}. The evolution law in the long-wavelength limit is,
\begin{equation}
\frac{\partial \eta_j}{\partial t} =-|\mathbf{k}_j|^2 \frac{\delta F}{\delta \eta_j^*},
\label{eq:amplitudetime}
\end{equation}
with
\begin{equation}
\begin{split}
\frac{\delta F}{\delta \eta_j^*}=& \left[
\Delta B_0 + B_0^x\mathcal{G}_j^2 + 3v \left(A^2-|\eta_j |^2\right)\right]\eta_j \\ 
& + \frac{\delta f^s(\{\eta_j\},\{\eta^*_j\})}{\delta \eta_j^*} .
\end{split}
\label{eq:amptimefuncder}
\end{equation}

In an equilibrium crystalline state, $A^2$ is a constant independent of crystal orientation. Thus, it supplies information about the order of the crystal phase. In particular, it has the maximum value in the relaxed crystal, decreases at defects and solid-liquid interfaces and vanishes in the disordered or liquid phase. For bulk crystals, the amplitude functions are constant. By assuming the amplitudes to be real and equal, i.e., $\eta_j=\phi_{0}$, it is possible to determine $\phi_{0}$ by minimizing the free energy in Eq.~\eqref{eq:energyamplitude}. The assumption of equal amplitudes holds true for triangular-honeycomb and bcc symmetries. For the fcc symmetry, the amplitudes are found to have different values depending on the magnitude of $\mathbf{k}_j$, i.e., they can be written as $\eta_j=\phi_{0,j}=\xi$ for $j \le 4$ and $\eta_j=\phi_{0,j}=\psi$ for $j\ge 5$, since $|\mathbf{k}_{j\ge 5}| = 2/\sqrt{3}|\mathbf{k}_{j\le 4}|$. 
Details about calculating $\phi_0$, $\xi$ and $\psi$ according to the selected lattice symmetry are reported in Appendix \ref{app:A} (hereafter we just use $\phi_0$ to denote $\phi_{0,j}$). 

When rotated or strained crystals are considered, the $\eta_j$'s become complex functions. For instance, the amplitude complex functions of a crystal phase rotated by an angle $\theta$ about z-axis are given by 
\begin{equation}
\eta_j = \phi_0\, e^{i\delta\mathbf{k}_{j}(\theta) \cdot \mathbf{r}},
\label{eq:amprot}
\end{equation}
where 
\begin{equation}
\begin{split}
\delta\mathbf{k}_{j}(\theta) = & \left[k^x_{j} (\cos\theta -1) - k^y_{j}
\sin\theta\right]\hat{\mathbf{x}} \\ &+ \left[k^x_{j} \sin\theta  + k^y_{j}
(\cos\theta-1)\right]\hat{\mathbf{y}} .
\end{split}
\label{eq:krot}
\end{equation}
On the other hand, a strained crystal can be described by 
\begin{equation}
\eta_j=\phi_0 \, e^{i\mathbf{k}_j \cdot \mathbf{u}(\mathbf{r})},
\label{eq:etaelas}
\end{equation}
where $\mathbf{u}(\mathbf{r})$ corresponds to the displacement field with respect to the relaxed crystal. Eq.~\eqref{eq:amprot} and \eqref{eq:etaelas} will be used in order to set the initial conditions for stressed and rotated crystals.

\subsection{Additional energy term}
\label{sec:coreenergy}

In Ref.~\cite{Guo2016} a term to control the interfacial free energy was introduced in the PFC model. This was achieved by considering a contribution to the free energy as $|\nabla \widetilde{n} |^2$, where $\widetilde{n}=\int d\mathbf{r} \chi(\mathbf{r}-\mathbf{r}')n(\mathbf{r})$ and $ \chi(\mathbf{r}-\mathbf{r}')$ is a smoothing function \cite{Kocher2015}, chosen to select density contributions on long wavelengths.
With this choice, variations of the density on short length scales are filtered out, while the ones present at the interfaces between phases remain, i.e. $\widetilde{n}$ is constant within bulk regions and changes
only at solid-liquid interfaces. Although not addressed in \cite{Guo2016}, this term would also impact the energy of dislocations or any defects in the crystal since the density is typically lower near such regions.

\begin{figure*}
\center
    \includegraphics[width=\linewidth]{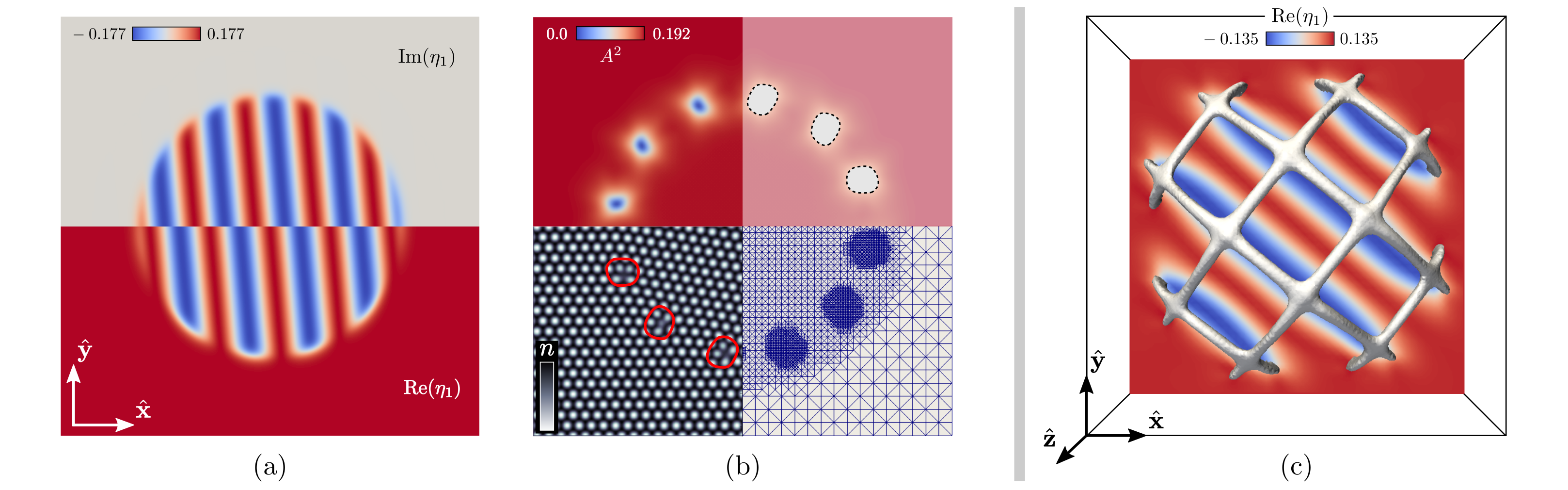} \caption{Illustrative application of the APFC model in two dimensions (a,b) and three dimensions (c). (a) Definition of a spherical tilted crystal with triangular lattice symmetry embedded in a relaxed crystal with the same crystal structure. The amplitudes, here
illustrated by means of the real (bottom) and imaginary (top) part of $\eta_1$, are constant and real in the surrounding relaxed crystal, while they oscillate in the embedded tilted crystal.  (b) Various visualizations (clockwise from the upper left panel):  $A^2$, the definition of defects according to a threshold of $A^2$, mesh refinement and reconstruction of the crystalline structure. (c) Dislocations network for a 3D rotated spherical grain in an fcc crystal superimposed to a central slice of the simulation domain showing the real part of $\eta_1$. Details are given in the main text.}
    \label{fig:figure1}
\end{figure*}

	In this work, we consider the APFC model in the absence of an average density term. Thus, the approach proposed in Ref.~\cite{Guo2016} cannot be directly considered within our framework. However, similar information is directly gathered from $A^2$, which is a measure of the crystalline order, and from its variation in space. In order to control the energy of interfaces or defects, we thus focus on a term involving only the gradient of $A^2$. In particular, in analogy with the gradient term in interfacial free energies \cite{Cahn1958}, we introduce the following additional energy contribution in Eq.\eqref{eq:energyamplitude}: 
\begin{equation}
F_{\beta}= \int_\Omega  \frac{\beta}{4}|{\nabla}A^2|^2 d\mathbf{r},
\label{eq:coreenergy}
\end{equation}
where $\beta$ is a free parameter. This leads to an additional term to Eq.~\eqref{eq:amptimefuncder} as
\begin{equation}
\frac{\delta F_{\beta}}{\delta \eta_j^*} = -\beta\eta_j \nabla^2 A^2.
\label{eq:evolcoreenergy}
\end{equation}
The additional energy term in Eq.~\eqref{eq:coreenergy} is then introduced phenomenologically. In the next sections, the influence of this term on the energy and morphology of interfaces and defects is investigated.

In the following, we refer to the total energy $F$ as the sum of the contributions in Eqs.~\eqref{eq:energyamplitude} and \eqref{eq:coreenergy}. As mentioned above, the specific form in Eq.~\eqref{eq:coreenergy} was chosen to modify the energy near dislocations and interfaces, but not to alter the elastic or other properties within bulk phases.  In this formulation to leading order, elastic strains in the system are incorporated in the phase of the complex amplitudes, of which $A^2$ is independent. For very large strains the magnitude of the complex amplitudes will be slightly altered and thus will alter $A^2$, but not $\nabla A^2$.  If the strain varies greatly over the sample Eq.~\eqref{eq:coreenergy} will be non-zero, but still small compared to the values near dislocation cores and interfaces.

\subsection{Numerical approach}
\label{sec:num}

A semi-implicit time discretization scheme is used in order to solve the set of equations defined in \eqref{eq:amplitudetime} and \eqref{eq:amptimefuncder}, and it is reported in detail in Appendix \ref{app:B1}. It consists of solving four second-order partial differential equations (PDEs) for each amplitude function. Different amplitudes are coupled due to the terms involving $f^s$ and $A^2$ in the evolution law, which are treated explicitly. For similar numerical approaches in solving PDEs for
materials-science applications, see, e.g., Refs.~\cite{Raetz2006,Salvalaglio2015}. 
The spatial discretization is done by FEM exploiting the adaptive finite-element toolbox AMDiS \cite{Vey2007,Witkowski2015}. We consider a refinement of the spatial discretization where the real and complex parts of $\eta_j$ oscillate, i.e., the regions where strained or tilted crystals are present.
Additionally, the refinement is increased at solid-liquid interfaces and defects, that is, where $A^2$ changes. Further details are given in Appendix \ref{app:B2}. Periodic boundary conditions (PBC) are considered for every simulation reported in the following. All the simulations are performed in parallel. 

Fig.~\ref{fig:figure1} shows sample simulations in two and three dimensions. In a relaxed crystal, a rotated spherical crystal of the same symmetry is embedded. In both cases, the initial configuration first forms a set of regular defects defining the GB. Then, in order to minimize the grain-boundary energy, the
embedded crystal begins to rotate and shrinks~\cite{Yeon2010}. Here we show snapshots when the defects are well defined and before much grain shrinkage occurred. 

In the 2D case, reported in Figs.~\ref{fig:figure1}(a) and \ref{fig:figure1}(b), the embedded crystal is rotated by 10$^\circ$ with respect to the surrounding matrix. Thus, the real and imaginary parts of the amplitudes vary in agreement with Eq.~\eqref{eq:amprot}. Even though the single amplitudes oscillate in the embedded crystal, $A^2$ is constant and only varies at the defects. The defects are located using a threshold for $A^2$ : $A^2<0.75\max(A^2)$ [see Fig.~\ref{fig:figure1}(b), upper part]. The computational grid is refined in the
embedded crystal due to the variation in $\eta_j$ and at the defects due to the variation in $A^2$ [see Fig.~\ref{fig:figure1}(b), lower right corner]. Reconstructing the density according to Eq.~\eqref{eq:amplitudetime} allows us to directly show the crystalline structure and identify the defects as illustrated in Fig.~\ref{fig:figure1}~(b), lower left corner. Solid red lines therein correspond to the $A^2=0.75\max(A^2)$ isolines.

The equivalent situation in three dimensions is shown in Fig.~~\ref{fig:figure1}(c) for a fcc crystal. The spherical GB is defined by a network of defects reflecting the cubic symmetry of the fcc crystal. Such defects are illustrated by means of the region where $A^2$ is below the threshold as in Fig.~\ref{fig:figure1}~(b, upper right corner). A central slice of the simulation domain is also shown, illustrating the oscillation of the real part of $\eta_1$ in the tilted crystal. A more detailed discussion of defect networks is given in Sec.~\ref{sec:strained}.

\section{Tuning the solid-liquid interfacial energy}
\label{sec:surf}

Let us consider a solid-liquid interface, where the solid is a relaxed crystal with $\eta_j=\phi_0$ in the bulk and $\eta_j=0$ in the liquid phase. Without loss of generality, we focus here on the 2D triangular symmetry for the crystalline solid phase. We consider the equilibrium condition at which the solid and the liquid phase have the same energy by setting $\Delta B_0=8t^2/(135v)$. As addressed in Ref.~\cite{Galenko2015}, by assuming real and identical amplitudes and focusing on the $\Delta B_0>0$ case, the equation describing the interface profile $\phi$ is in the long wavelength limit, 
\begin{equation}
2B^x\nabla^2 \phi-\Delta B_0\phi+2t\phi^2-15v\phi^3+6\beta \phi
\nabla^2 \phi^2=0,
\label{eq:mub0}
\end{equation}
which corresponds to a stationary interface. Moreover, the condition $\phi=\phi_0$ in the bulk crystal and $\phi=0$ in the liquid phase must be satisfied. For $\beta=0$ this can be solved analytically by assuming a tanh-profile for $\phi$ perpendicular to the solid-liquid interface:
 \begin{equation}
\phi=\frac{\phi_0}{2} \left[1-\tanh\left(\frac{x}{\chi} \right)\right].
\label{eq:tanh}
\end{equation}
Equation \eqref{eq:mub0} is then solved by
\begin{equation}
\chi=\frac{4}{\phi_0}\sqrt{\frac{B^x}{15v}}=\frac{3\sqrt{15vB^x}}{t},
\label{eq:pareq}
\end{equation}
where $\phi_0=4t/(45v)$ at equilibrium. For $\beta \neq 0$ this ansatz does not lead to a solution of Eq.~\eqref{eq:mub0}.  However, it is expected to properly describe the amplitude profile at the solid-liquid surface in the $\beta\rightarrow0$ limit \cite{Galenko2015}. Thus, in this limit, we can estimate the contribution due to $\beta$ assuming that $\phi$ is not significantly influenced by the additional energy term. For a straight interface the energy contribution due to $\beta$ from Eq.~\eqref{eq:coreenergy} is approximately,
\begin{equation}
\frac{\ell \beta}{4}\int_{-\infty}^{\infty} |{\nabla}A^2|^2 dx =
\frac{18\phi_0^4}{5\chi} \ell \beta,
\label{eq:approxene}
\end{equation}
where $\ell$ is the length of the interface. 

\begin{figure}
\center
    \includegraphics[width=\linewidth]{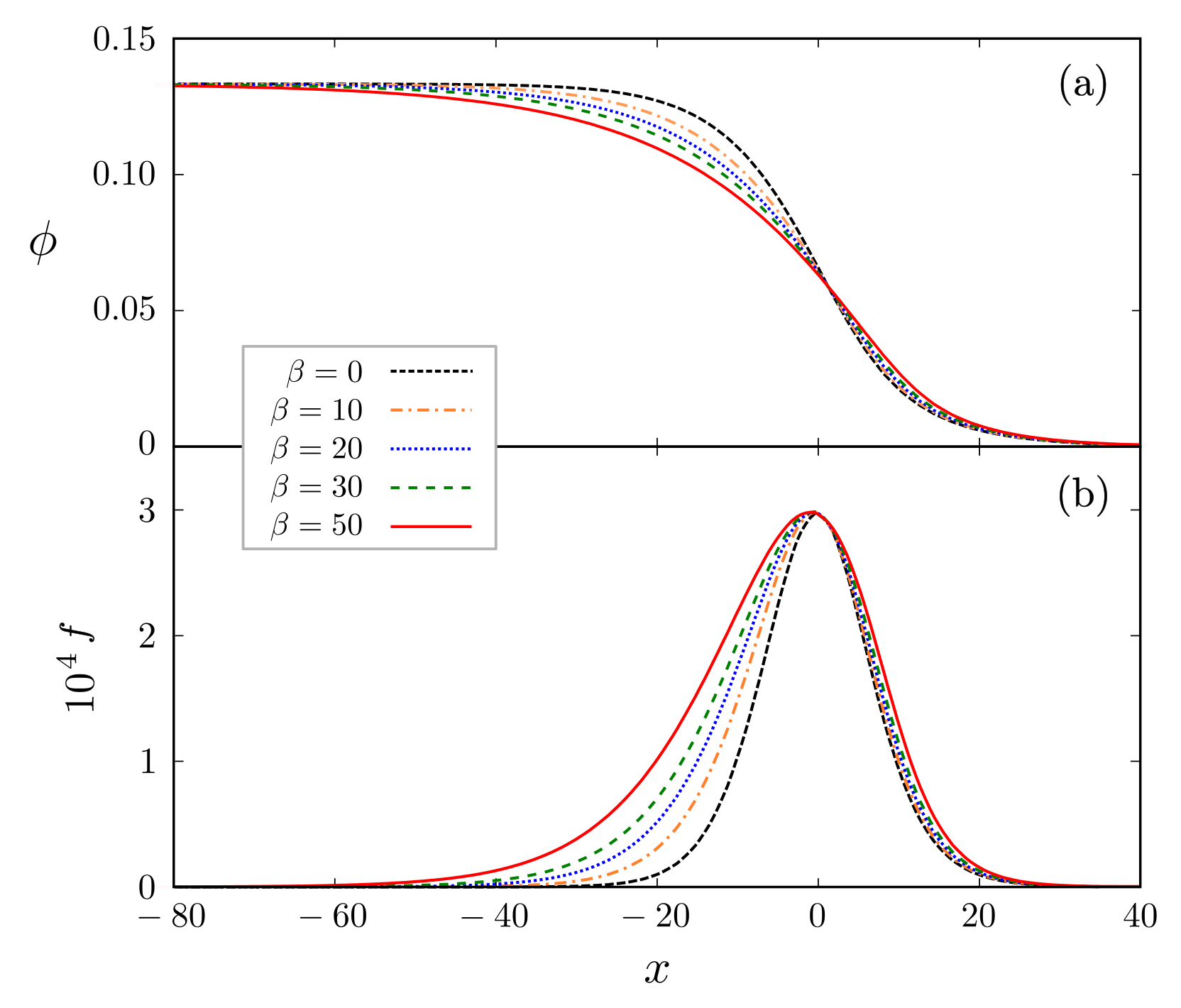} \caption{Effect of the additional energy term on the solid-liquid interface properties. (a) Profiles perpendicular to the solid-liquid interface at equilibrium in terms of $\phi=\sqrt{A^2/6}$ with $\beta \in [0; 50]$. (b) Energy density corresponding to the different profiles in panel (a).}
    \label{fig:figure2}
\end{figure}
To evaluate the contribution to the interfacial energy of the term in Eq.~\eqref{eq:coreenergy}, to show the change in the interface morphology, and to check the validity of the approximation in Eq.~\eqref{eq:approxene}, we solve the equations of the APFC model numerically for a straight, solid-liquid interface. The parameters are set as follows: $B^x=0.98$, $v=1/3$, $t=1/2$ and $\Delta B_0$ in order to achieve equilibrium condition. The results are shown in Fig.~\ref{fig:figure2}, which illustrates the effect of different $\beta$ values on the solid-liquid interface. In particular, Fig.~\ref{fig:figure2}(a) shows the profile perpendicular to the interface in terms of $\phi$, obtained as $\phi=\sqrt{A^2/6}$, which minimizes the energy for different $\beta$ values. For $\beta=0$ it is well described by the function in Eqs.~\eqref{eq:tanh} and \eqref{eq:pareq} in agreement with Ref.~\cite{Galenko2015}. By increasing $\beta$ the width of the interface increases. Moreover, the region closer to the solid phase undergoes a more significant smearing than the one close to the liquid phase. Thus, it does not qualitatively correspond to a tanh-profile as described in Eq.~\eqref{eq:tanh}. Fig.~\ref{fig:figure2}(b) illustrates the changes of the energy density $f$, such as $F=\int_\Omega f(\mathbf{r})d\mathbf{r}$, at the interface with increasing $\beta$. According to the modification of the interface profile, the region with an
energy density larger than zero increases for larger $\beta$, with a smaller gradient towards the solid phase. The maximum value of $f$ is found to be not significantly affected by the additional energy contribution and it shifts slightly towards the solid phase. 

\begin{figure}
\center
    \includegraphics[width=\linewidth]{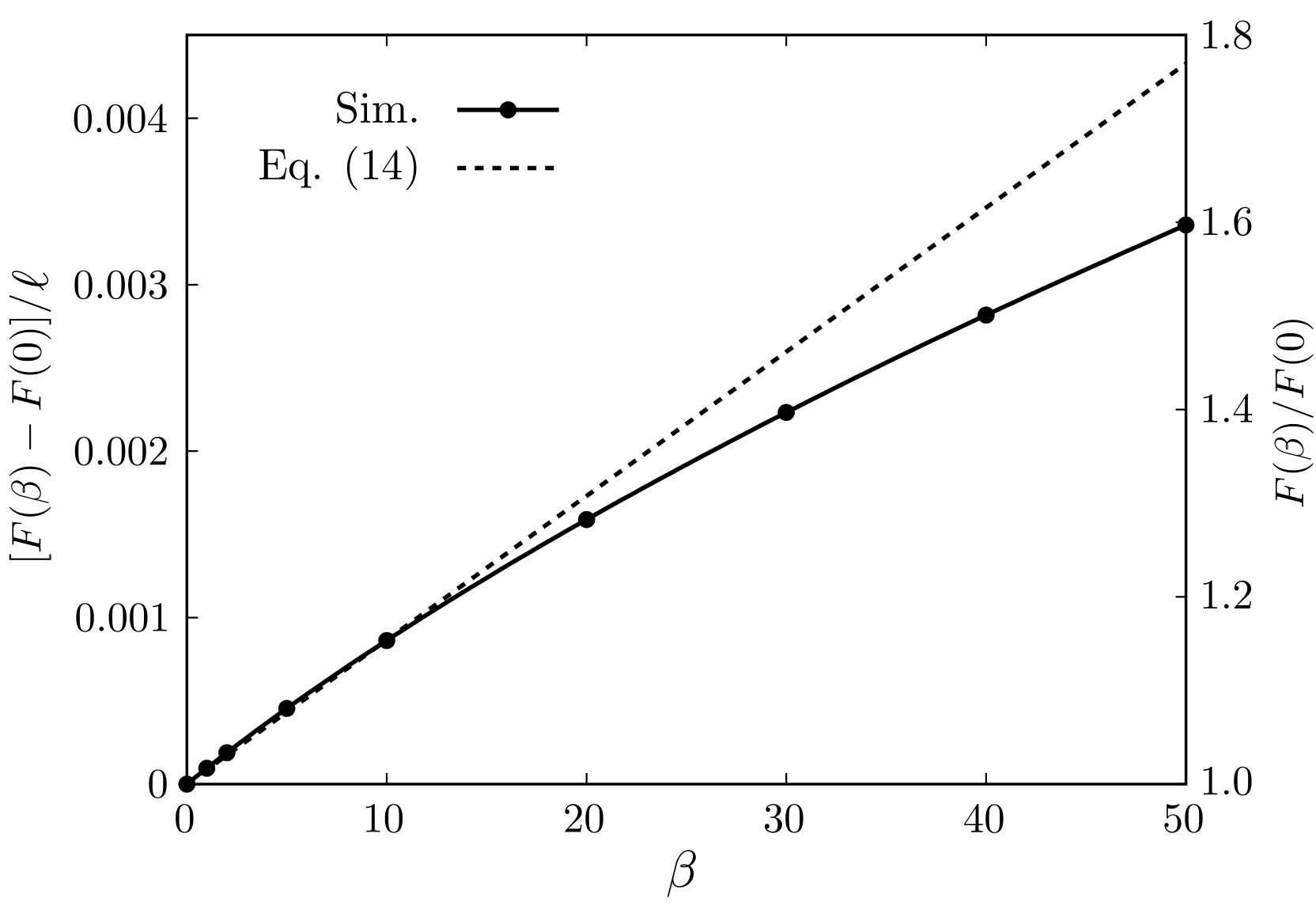} \caption{Excess of energy per unit length due to the presence of the interface as a function of $\beta$, $[F(\beta)-F(0)]/\ell$. The dots represents the simulations, shown here with a solid guideline. The dashed line represents the values predicted by neglecting the changes in the interface profile as in Eq.~\eqref{eq:approxene}. Additionally, the right y-axis shows the relative change in the surface energy $F(\beta)/F(0)$, with $F(0)=0.005574$.}
    \label{fig:figure3}
\end{figure}

In Fig.~\ref{fig:figure3} the change in the total energy due to $\beta$ is shown. For small $\beta$, the increase of the interface energy density is nearly linear and corresponds well with the approximation reported in Eq.~\eqref{eq:approxene}. In this case, the morphology of the profile of $\phi$ is not significantly altered and the assumption leading to Eq.~\eqref{eq:approxene} is well fulfilled. For larger $\beta$, more significant deviations are observed, providing a sub-linear behavior. 
Within the range of $\beta$'s used, a relative scaling factor of up to $\sim 1.6$ can be achieved, and no restrictions are present for larger values. This can be used in order to match the solid-liquid interface energies from experiments or first-principles approaches, while they are typically underestimated in classical PFC methods \cite{Guo2016}. Negative values of $\beta$, even small ones, lead to instabilities in the solid phase. This restricts $\beta$ to be positive in practice.

\section{Tuning the energy of defects between tilted crystals}
\label{sec:tilted}

\begin{figure}
\center
    \includegraphics[width=\linewidth]{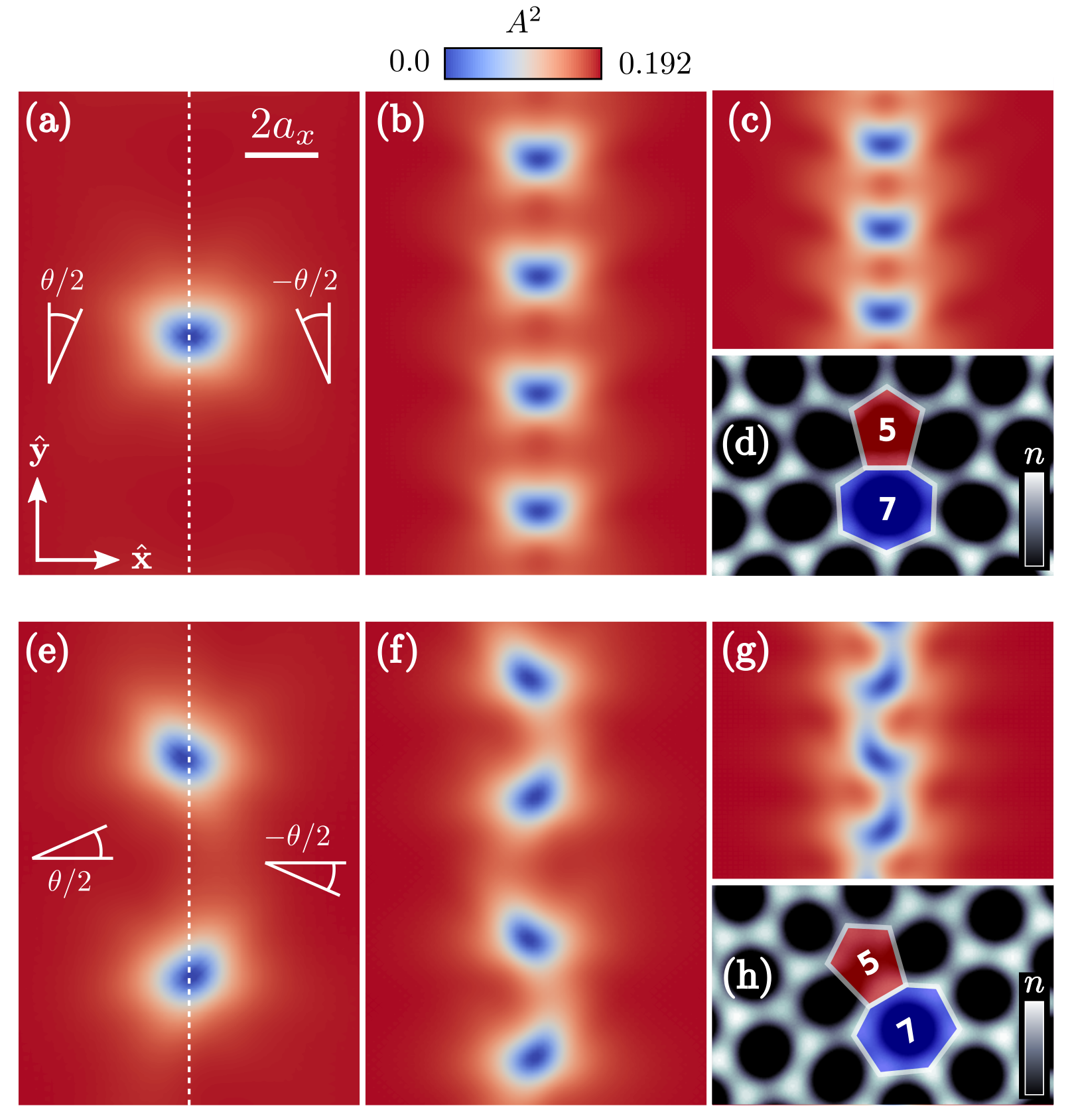} \caption{Dislocations forming at the grain boundaries. (a) $\theta=6.3^\circ$, (b) $\theta=18.8^\circ$, (c) $\theta=26.3^\circ$, for vertical GBs. (d) Magnification of the defect showing a graphene-like, continuous density $n$ as obtained from Eq.~\eqref{eq:density} using the amplitudes in panel (a). (e) $\bar{\theta}=7.8^\circ$, (f) $\bar{\theta}=14.7^\circ$, (g) $\bar{\theta}=26.3^\circ$, for horizontal GBs after rotation by $90^\circ$. (h) As in panel (d) the $n$ is reconstructed using the amplitudes in panel (e). The $5|7$ structure of defects is illustrated in panels (d) and (h). $a_x=4\pi/\sqrt{3}$.}
\label{fig:figure4}
\end{figure}

In this section, we describe the effect of the additional energy term in Eq.~\eqref{eq:coreenergy} on the morphology of defects occurring between tilted crystals and on their energy. In particular, the relevant case of the 2D honeycomb structure is considered \cite{Hirvonen2016}. The parameters are set as in the previous section, with $\Delta B=0.02$ and $t=-1/2$ for which the equilibrium state is a honeycomb crystalline phase.
A rectangular domain, $L_x \times L_y$ with $\hat{\mathbf{x}}=[10]$ and $\hat{\mathbf{y}}=[01]$, is considered with a straight vertical GB at the center, forming between two 2D tilted crystals. The relative tilt angle between the two crystals, $\theta$, is set by initializing the $\eta_j$ functions with Eqs.~\eqref{eq:amprot} and \eqref{eq:krot} and imposing a $\pm \theta/2$ tilt for the left and the right part of the simulation domain respectively, as also illustrated in
Fig.~\ref{fig:figure4}(a). By using PBC, a GB with infinite extension is considered. Moreover, a second GB is expected, which is shared between the left and right boundary of the simulation domain. $L_x$, (twice the distance between GBs along $\hat{\mathbf{x}}$ direction) can be chosen arbitrarily and it is set here to be significantly larger than the spacing of the defects at the GB. Additionally, care has to be taken in choosing $L_y$, so that the periodicity of amplitudes along the $\hat{\mathbf{y}}$ direction fit the domain. The details about choosing $\theta$ and the domain size in order to ensure this condition are summarized in Appendix \ref{app:B3}.

The APFC approach well describes GBs for small $\theta$. For large tilts, it does not predict their correct morphologies \cite{Spatschek2010}. However, the GB obtained for large $\theta$ can be simulated by considering a similar tilt as before, called here $\bar{\theta}$, but with a horizontal GB. Therefore, $\pm \bar{\theta}/2$ are set in the top and bottom region of the rectangular domain (as shown in Fig.~\ref{fig:figure4}(e) after a rotation of the domain by $90^\circ$). The results with the two configurations can then be compared considering $\theta=60^\circ-\bar{\theta}$. In this case, $L_y$ is chosen larger than the spacing between defects, and $L_x$ is set as described in Appendix \ref{app:B3}.

Let us consider first the $\beta=0$ case. Fig.~\ref{fig:figure4}(a)-(c) show the defects at the vertical GB for different values of $\theta$ by means of $A^2$. The larger the tilt is, the larger is the density of dislocations along the GB. The morphology of defects is similar for the different cases, but for large tilts their superposition increases. Figs.~\ref{fig:figure4}(e)-(g) show the similar behavior obtained by increasing $\bar{\theta}$ for the horizontal GB and rotating the domain by $90^\circ$ in order to provide a better comparison to the aforementioned case.  The angular dependence of dislocation density is similar, but the arrangement of the defects is different.

	To more closely examine the dislocations, amplitude functions can be used to reconstruct the density by means of Eq.~\eqref{eq:density} for a honeycomb lattice. This is done in Figs.~\ref{fig:figure4}(d) and \ref{fig:figure4}(h) for a vertical and horizontal GB as discussed before. The two different grain boundaries observed for these structures, namely the armchair (AC) GB in \ref{fig:figure4}(a)-(d) and the zigzag (ZZ) GB in  \ref{fig:figure4}(e)-(h) are observed. As highlighted in the corresponding figures, both cases are compatible with the peculiar $5|7$ arrangement of atoms at the defects between tilted graphene layers \cite{Hirvonen2016}.

\begin{figure}
\center
    \includegraphics[width=\linewidth]{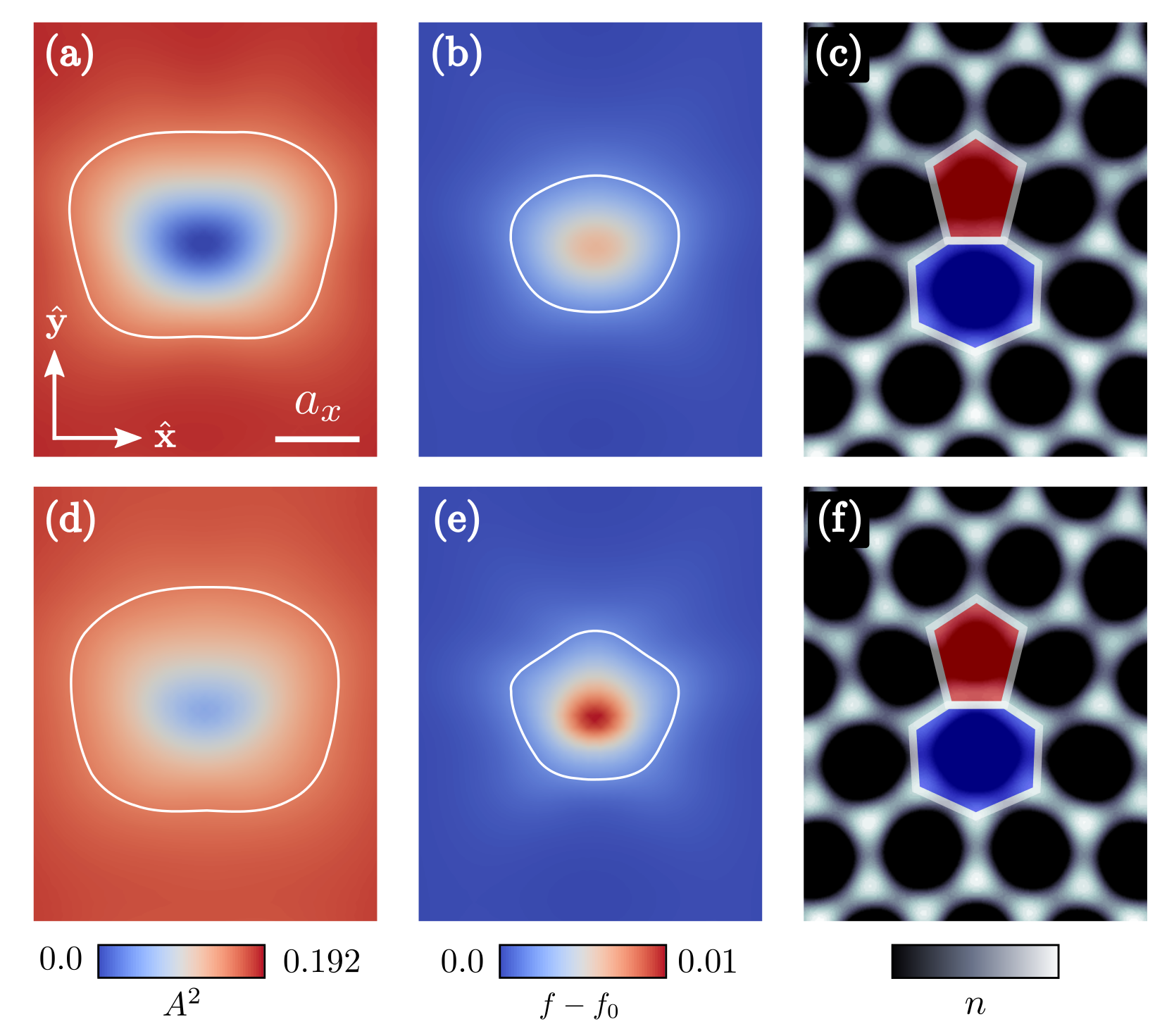} \caption{Effect of the core-energy term on the defect features. (a) and (b) show the values of $A^2$
and $f-f_0$ respectively for a dislocation forming at a AC-GB with $\theta = 12.8^\circ$ with $\beta=0$. (d) and (e) show $A^2$ and $f-f_0$
respectively for the same dislocation as in (a) and (b) with $\beta=10$. Isolines are also shown corresponding to $A^2=0.16$ in panel (a) and (d) and
$f-f_0=10^{-5}$ in panel (b) and (e). (c) and (f) show $n$ at the defects with $\beta=0$ and $\beta=10$ respectively.} 
\label{fig:figure5}
\end{figure}

The effect of non-zero $\beta$ values is shown in Fig.~\ref{fig:figure5}. Here a single dislocation at a vertical GB with $\theta=12.8^\circ$ is highlighted. Fig.~\ref{fig:figure5}(a) shows $A^2$ at the defect for $\beta=0$. Fig.~\ref{fig:figure5}(b) shows the excess of the energy density with respect to the bulk crystal, $f-f_0$, for such a defect. The same quantities are shown in Fig.~\ref{fig:figure5}(d) and (e) for $\beta=10$. The change in the $A^2$ field can be easily noticed. The depth of the minimum decreases with increasing $\beta$, while the energy density increases with increasing $\beta$. Despite these changes, the reconstructed density as shown in Fig.~\ref{fig:figure5}(c) and (f), remains unaltered. This is mainly due to the fact that the extension of the region where $A^2$ decreases and $f-f_0$ is larger than zero does not change significantly by increasing $\beta$ (see the solid, white isolines).  According to these results, the effect of the additional energy term consists of an increase of the energy at the defect, without affecting the type of defect and the corresponding arrangement of atoms in the crystal lattice.

\begin{figure}
\center
    \includegraphics[width=\linewidth]{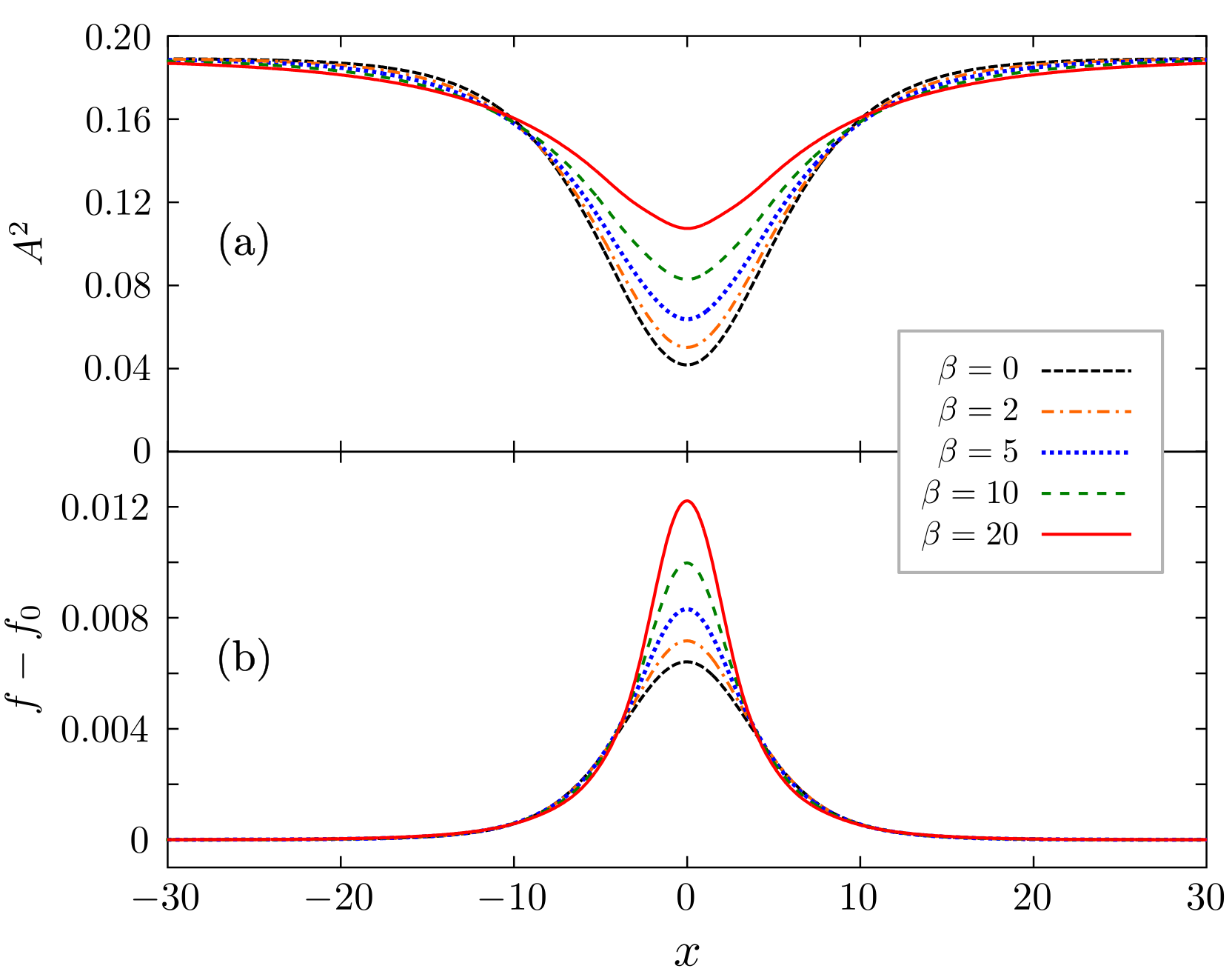} \caption{Line scans along the $\hat{\mathbf{x}}$ direction passing through the center of the defect in Fig.~\ref{fig:figure5} (i.e. the minimum of $A^2$) showing (a) $A^2$ and (b) $f-f_0$, for different $\beta$ values.}
    \label{fig:figure6}
\end{figure}

A more quantitative comparison is performed in Fig.~\ref{fig:figure6}, which shows $A^2$ and $f-f_0$ along a horizontal line passing through the center of the defect and perpendicular to the straight GB line to which it belongs. In particular, the order parameter $A^2$ is slightly broader for larger $\beta$ values as illustrated in Fig.~\ref{fig:figure6}(a). However, Fig.~\ref{fig:figure6}(b) shows that the additional energy contribution controlled by $\beta$ is localized at the defects. Indeed, it affects only the maximum at the center of the defect, while it decreases when moving away from the GB with a decay rate nearly independent on $\beta$. 

\subsection{Control of grain boundary energy}

\begin{figure}
\center
  \includegraphics[width=\linewidth]{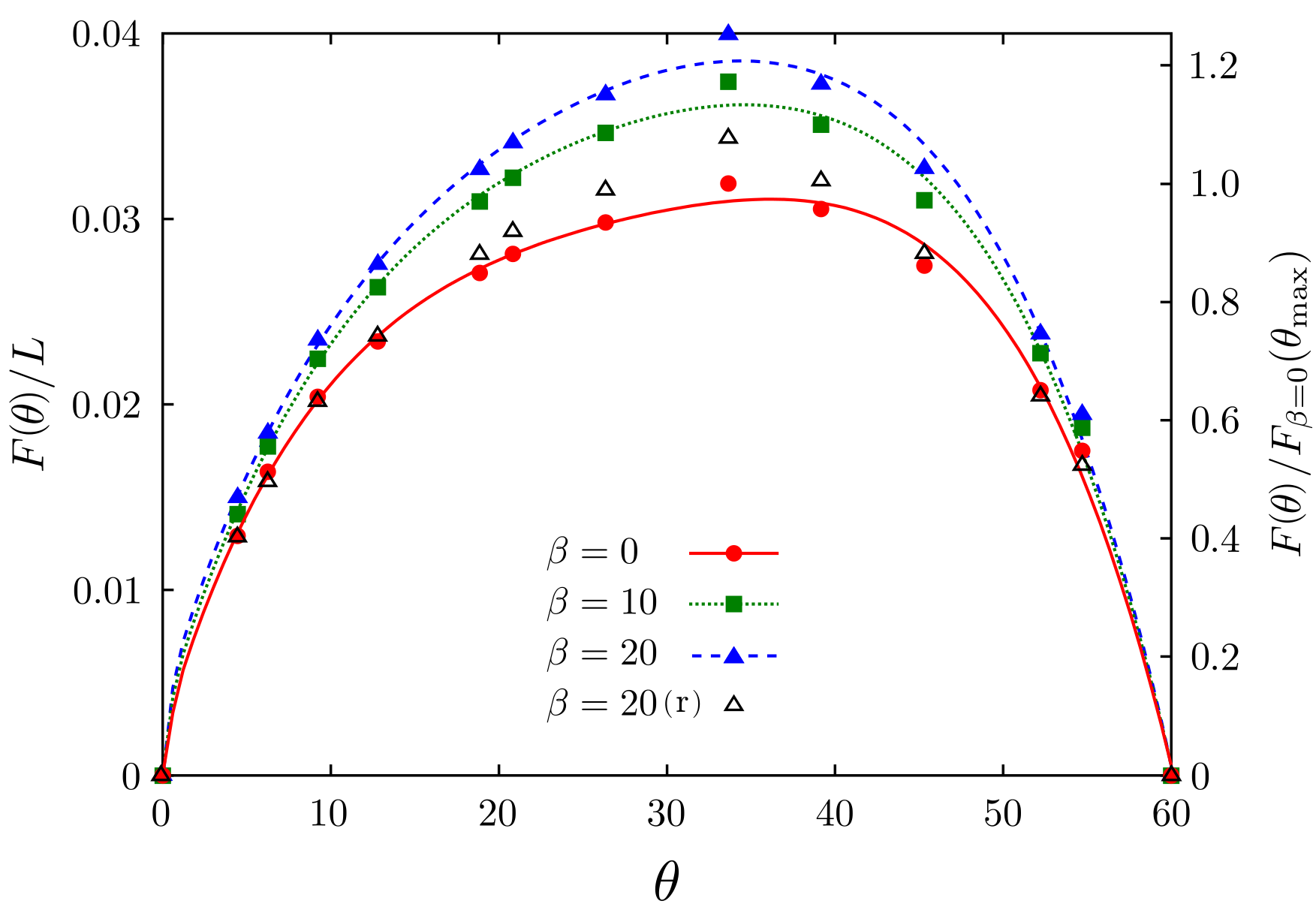} \caption{Grain boundary energy as function of the tilt angle for different $\beta$ values. The energy per unit length, $F(\theta)/L$, and the normalized energy with respect to the maximum energy value for the $\beta=0$ case, $F(\theta)/F_{\beta=0}(\theta_{\text{max}})$, are shown. Energy values for GBs with $\theta<30^\circ$ are obtained with the AC-GB configuration. Values with $\theta>30^\circ$ are obtained with the ZZ-GB configuration with tilt angle $\bar{\theta}$ and $\theta=60^\circ-\bar{\theta}$. Interpolated guidelines are superimposed to the symbols corresponding to the results of simulations: $\beta=0$ (red dots, solid guideline), $\beta=10$ (green squares, dotted guideline), $\beta=20$ (blue triangles, dashed guideline). Empty triangles correspond to the energy values of the $\beta=20$ case, rescaled (r) in order to have the same value at $\theta \approx 4.45^\circ$ as with $\beta=0$.}
  \label{fig:figure7}
\end{figure}

After assessing the role of the additional energy term on the features of a single dislocation, the global effect when looking at the energetics of a GB as a whole can be considered. In particular, we focus here on the dependence of the energy per unit length of GBs, $F/L$, as a function of the tilt angle $\theta$.  Within our framework we can simulate all the possible angles by considering vertical GBs for $\theta<30^\circ$, and horizontal GBs for $\theta > 30^\circ$ by setting $\bar{\theta}<30^\circ$ and $\theta=60^\circ-\bar{\theta}$ \cite{Hirvonen2016}. In Fig.~\ref{fig:figure7} we report the energy per unit length of GBs in terms of $F(\theta)/L$ and $F(\theta)/F_{\beta=0}(\theta_{\text{max}})$. The latter corresponds to a normalization of the calculated energies with respect to the maximum value of the $\beta=0$ curve. The red dots correspond to the results obtained with $\beta=0$. A solid guideline is also superimposed to the simulation results, reproducing the typical energy dependence on the tilt expected for these systems. Such a result directly corresponds to what is obtained in Ref.~\cite{Hirvonen2016}, further assessing our computational approach. The simulation results obtained by considering $\beta=10$ and $\beta=20$ are also shown by green squares and blue triangles together with dotted and dashed guidelines, respectively. They reveal the global effect of the new energy term on the $F(\theta)/L$ curves. The increase of defect energy due to $\beta$ also leads to an overall increase of the GB energy. For instance, a relative increase of a factor $\sim1.25$ is obtained for the $\beta=20$ case for the maximum of the energy. This relative change is similar to what was obtained in the tuning of the solid-liquid interfacial energy (see Fig.~\ref{fig:figure3}). More detailed insights can be obtained by considering 
the empty triangles shown in Fig.~\ref{fig:figure7}, which correspond to the $F_{\beta=20}(\theta)/L$ curve rescaled in order to have the same value at $\theta=\theta^* \approx 4.45^\circ$ with $\beta=0$ case, i.e. multiplied by $F_{\beta=0}(\theta^*)$/$F_{\beta=20}(\theta^*)$.  These values highlight the fact that a small change in the shape of the $F(\theta)/L$ curves is induced when considering nonvanishing $\beta$ values. That is, these curves are not self-similar.  The reason for this is that the higher angle GBs contain more dislocations and in turn more dislocation energy. Thus the higher angle GB energy increases more than the lower-angle GB energy when $\beta$ 
is increased. It is worth mentioning, however, that the observed change in the shape of the energy curves is in the order of the typical experimental fluctuation (see for instance the comparison between PFC calculations and experiments in Ref.~\cite{Elder2004}). Therefore, the increase of the energy obtained for a specific $\theta$ can be considered as representative of the effect on the entire $F(\theta)/L$ curve.

\section{Tuning the energy of defects in strained systems}
\label{sec:strained}
So far we investigated the case of defects when forming between tilted crystals. However, dislocations are known to form also when applying an external load to the material or at the interface between mismatched, epitaxial structures in order to relive the resulting stress and lower the elastic energy. In this section we consider 2D and 3D multilayer structures where subsequent layers have opposite in-plane strain $\pm \varepsilon$. When considering a 2D system,
$L_x \times L_y$ as in Sect.~\ref{sec:tilted}, with PBC and the normal to the interface between layers along the $\hat{\mathbf{y}}$ direction, the configuration can be initialized using Eq.~\eqref{eq:etaelas} with the following displacements, 
\begin{equation}
\mathbf{u}(\mathbf{r}) = \left\{ 
\begin{split}
-u_x \hat{\mathbf{x}}& \qquad    \frac{L_y}{2} < y < \frac{3 L_y}{4}  \\
+u_x \hat{\mathbf{x}}& \qquad    \text{elsewhere}  
\end{split}
\right.
\label{eq:initelas2D}
\end{equation}
with $u_x$=$a_x x/ L_x$, and $a_x$ is the distance between maxima of the density as in Eq.~\eqref{eq:density} along the $\hat{\mathbf{x}}$ direction. With this choice
$\varepsilon=\pm a_x / L_x$ and matching amplitudes are obtained at the boundaries. For 3D systems, $L_x \times L_y \times L_z$ with $\hat{\mathbf{x}}=[100]$, $\hat{\mathbf{y}}=[010]$, $\hat{\mathbf{z}}=[001]$, and normal to the interface between layers along the $\hat{\mathbf{z}}$ direction, the in-plane strain can be set as \begin{equation}
\mathbf{u}(\mathbf{r}) = \left\{
\begin{split}
-u_x \hat{\mathbf{x}}-u_y \hat{\mathbf{y}}& \qquad \frac{L_z}{2} < z < \frac{3 L_z}{4}  \\
+u_x \hat{\mathbf{x}}+u_y \hat{\mathbf{y}}& \qquad \text{elsewhere}  
\end{split}
\right.
\label{eq:initelas3D}
\end{equation}
with $u_x$ as in Eq.~\eqref{eq:initelas2D} and $u_y$=$a_y y/ L_y$ with $a_y$ the distance between maxima of the density \eqref{eq:density} along the $\hat{\mathbf{y}}$ direction. In this case $\varepsilon_x=\pm a_x / L_x$ and $\varepsilon_y=\pm a_y / L_y$. The parameters defined in Eq.~\eqref{eq:density}
are set as in Sec.~\ref{sec:tilted}.

\begin{figure*}
\center
  \includegraphics[width=\linewidth]{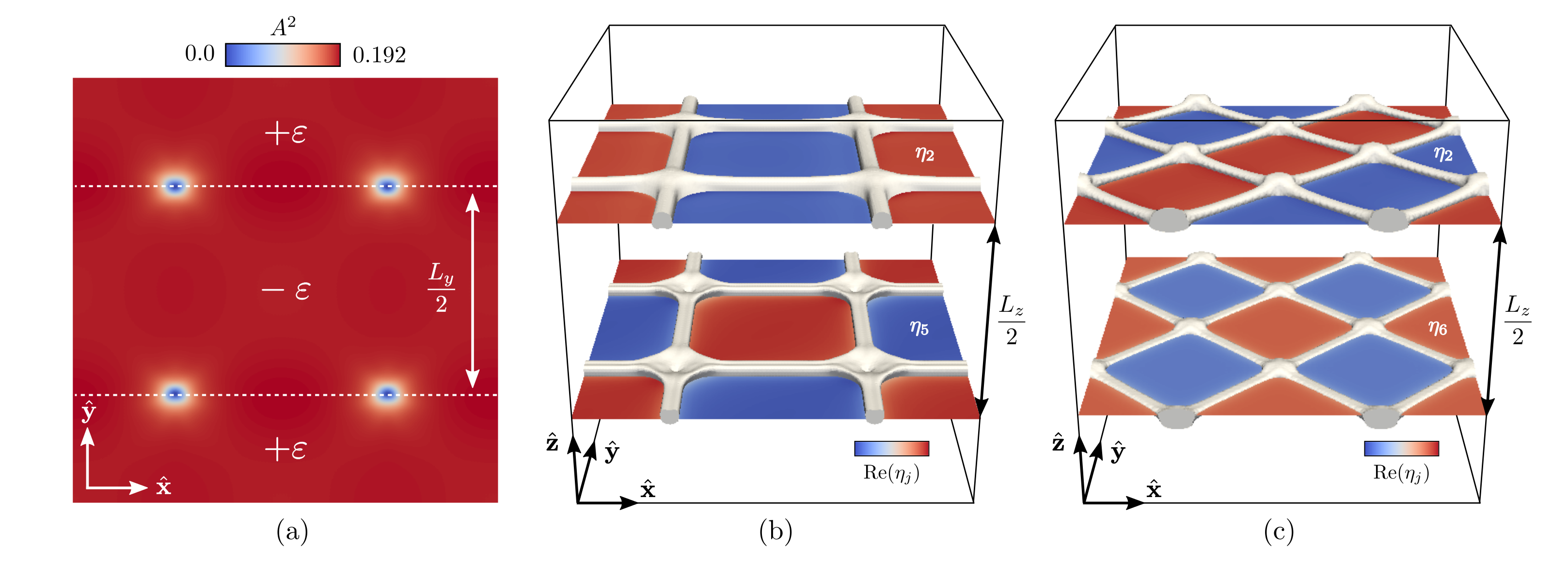} \caption{Defects in multilayer systems with alternate in-plane strain $\pm \varepsilon$. (a) Dislocations forming in a 2D crystal with triangular symmetry as resulting from the relaxation of the initial condition from Eq.~\eqref{eq:initelas2D}. (b) and (c) show dislocations as resulting from the relaxation of initial condition set by Eq.~\eqref{eq:initelas3D} for 3D crystals with a bcc and fcc lattice respectively. Dislocations in 3D are shown as in Fig.~\ref{fig:figure1}(c). The different colors in the planar regions bounded by dislocations in panel (b) and (c) illustrates the behavior of some representative amplitude functions at the interface between different layers (color online). All the panels show the defects when a stationary shape is obtained.} 
\label{fig:figure8}
\end{figure*}

The case corresponding to a triangular or honeycomb structure as in Eq.~\eqref{eq:initelas2D} is shown in Fig.~\ref{fig:figure8}(a). A square simulation domain is considered with $L_x=80\pi$. According to the definition of $\mathbf{k}_j$ vectors in \eqref{eq:ktri}, $a_x=4\pi/\sqrt{3}$. Then, a strain of $\varepsilon \approx \pm 0.029$ is applied in the two layers respectively. Notice that with this initial condition a difference of two lattice spacings $a_x$ is achieved across the interface. As a result of the evolution laws in Eq.~\eqref{eq:amplitudetime}, this initial condition evolves to two pairs of dislocations as depicted in Fig.~\ref{fig:figure8}(a). Despite the symmetric initial condition for the strain, the defects start to move after being formed. This is due to the asymmetry of the energy when considering opposite strain, leading to higher values when compressing the materials as it naturally accounts for repulsive effects when shortening the distance between atoms \cite{Emmerich2012}. Indeed, the motion of defects occurs in order to shrink the layer with negative strain. However, with the selected strain the motion after the formation of the defect is very slow, involving a timescale significantly larger than the formation of the defects from the considered initial condition. Fig.~\ref{fig:figure8}(a) corresponds to the stage at which the shape of the defect become stationary.

A similar configuration involving strained layers in 3D with bcc crystal symmetry is shown in Fig.~\ref{fig:figure8}(b). We consider a strained system as set by Eq.~\eqref{eq:initelas3D}. The periodicities of the atomic density, according to $\mathbf{k}_j$ vectors in \eqref{eq:kbcc} read $a_x=a_y=2\pi\sqrt{2}$. A cubic simulation domain is set with $L_x$=$80\pi$. With this choice $\varepsilon_x=\varepsilon_y \approx \pm 0.035$. The resulting dislocation network forming from the evolution of amplitudes at the interface between layers is shown in Fig.~\ref{fig:figure8}(b). In particular, the dislocation network is shown by means of $A^2$ values as in Fig.~\ref{fig:figure1}(c). The two interfaces between layers with opposite strain are shown by xy-planes, illustrating also the real part of two representative amplitude functions. In materials with this structure, dislocations are known to occur mainly with a $\{$110$\}\langle$111$\rangle$ slip system, and more rarely with a $\{$112$\}\langle$111$\rangle$ slip system \cite{HullBook}\footnote{As slip system we refer to a family of planes along which the dislocation may glide and a family of directions which correspond to magnitude and direction of the lattice distortion induced by the dislocation, i.e to the so-called Burgers vector.}. For instance, a prominent example consists of Fe crystals \cite{Queyreau2011}. As a result of the simulation approach considered here, dislocations form along $\hat{\mathbf{x}}$ and $\hat{\mathbf{y}}$ direction, which is compatible with the constraint of lying on $\{$110$\}$ planes (e.g. the (101) plane), from the slip system, and on the (001) interface, as it is the interface between layers with different strain from the initial condition. The cross-section of the defects aligned along the horizontal axis shows a structure similar to what is observed in Fig.~\ref{fig:figure8}(a). Also in this case, the structure in Fig.~\ref{fig:figure8}(b) refers to the stage where the shape of the dislocation network is stationary.

Fig.~\ref{fig:figure8}(c) shows the stationary shape resulting from a setup as in Fig.~\ref{fig:figure8}(b) with fcc crystal symmetry. Notice that this corresponds to a prototypical system for fcc materials \cite{Karnthaler78}, and it shows also similarities with technology-relevant zincblende or diamond structures \cite{Fitzgerald1991}. For this symmetry $a_x=a_y=2\pi \sqrt{3}$ as from Eqs.~\eqref{eq:kfcc}. A cubic simulation domain is set with $L_x=80\pi$. The resulting in-plane strain is then $\varepsilon_x=\varepsilon_y\approx \pm 0.043$. Starting from this initial condition, the evolution laws lead also in this case to the formation of a dislocation network at the interface. For dislocations in fcc crystals, a $\{$111$\}\langle$110$\rangle$ slip system is expected \cite{HullBook}. Dislocations are actually found to be aligned along the $\langle110 \rangle$ directions, which correspond to the intersections between some $\{$111$\}$ and (001) planes, i.e. to slip planes in fcc crystals and the interface between domains with different strains. Notice that the amplitudes values at the interface between layers illustrated in Fig.~\ref{fig:figure8}(c) show different maximum and minimum values. Indeed, they belong to the two groups of equivalent amplitudes playing a different role in the energy functional and having different values also when considering real, constant amplitudes in relaxed crystals (see also Appendix \ref{app:A}).

\begin{figure}
\center
  \includegraphics[width=\linewidth]{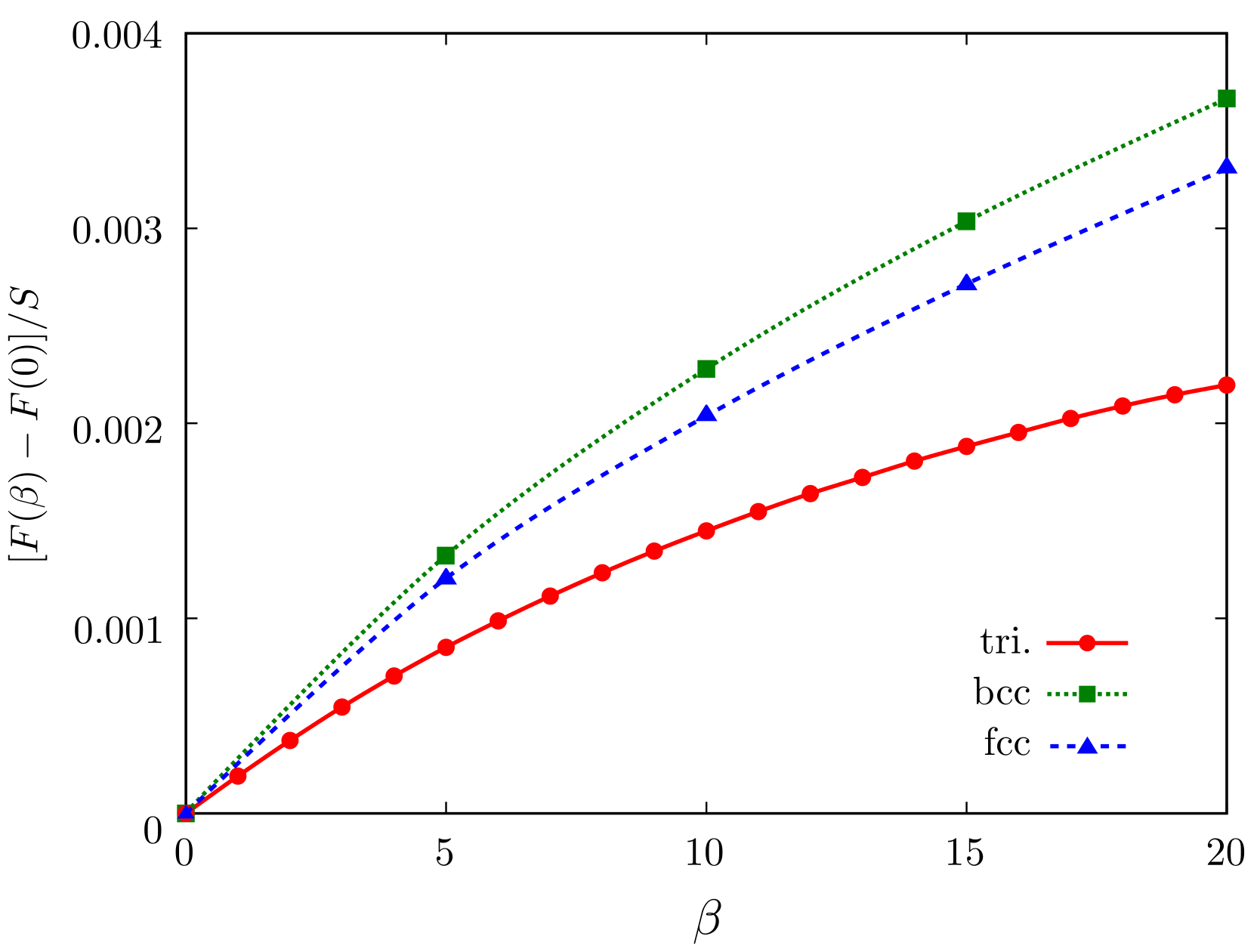} \caption{Excess of energy induced by the energy term of Eq.~\eqref{eq:coreenergy} when considering defects in strained systems as in Fig.~\ref{fig:figure8}. Different curves show such an effect for 2D triangular (red dots and solid guideline), 3D bcc (green squares and dotted guideline) and 3D fcc (blue triangles and dashed guideline) crystals. } \label{fig:figure9}
\end{figure}

Tuning of the energy for the 2D and 3D configurations reported in Fig.~\ref{fig:figure8} is shown in Fig.~\ref{fig:figure9}. In this plot we consider the difference in the total energy per interface length or area induced by the additional energy term of Sec.~\ref{sec:coreenergy}, namely $[F(\beta)-F(0)]/S$, where $S=L_x$ in 2D and $S=L_xL_y$ in 3D. As already observed in the previous sections, a linear dependence on $\beta$ of the energy increase is achieved for small $\beta$. Then a sublinear behavior is observed for larger $\beta$ values. By comparing the different symmetries, we can notice that a higher effect is achieved when considering 3D crystals. This may be ascribed to a denser configuration of defects, as a result of a biaxial strain in three dimensions instead of a uniaxial strain in two dimensions. This has been already observed in Fig.~\ref{fig:figure7} where the effect of $\beta$ is higher when increasing the tilt between the crystal, producing a larger number of defects per unit length (see also Fig.~\ref{fig:figure4}). Moreover, the bcc lattice shows a larger energy increase with $\beta$ than the fcc lattice, which can be ascribed to the larger region involving changes of $A^2$, i.e. the defects appears broader, as can be noticed from Fig.~\ref{fig:figure8}. We verified that for the results reported in this section, the changes in the defect morphology induced by the additional energy term in Eq.~\eqref{eq:coreenergy} are analogous to what is discussed in Sec.~\ref{sec:tilted}.

\section{Conclusions and Remarks}
\label{sec:conclusions}

 In this work, we extended the APFC model in order to tune the solid-liquid interface and defect energies, increasing the capabilities of the approach in the description of real material properties. 
 
 The effect of the additional energy term introduced in Sect.~\ref{sec:coreenergy} on the interface morphology as well as the increasing of the energy as a function of the control parameter $\beta$ were illustrated. Moreover, an approximate analytical expression was derived for the influence of $\beta$ on the solid-liquid interface, showing that for small $\beta$ the energy increase was linear in $\beta$.  
 
 The ability to tune the energy of defects at the GB between tilted crystals was then examined. The additional energy contribution is found to affect the minimum value of the order parameter $A^2$ at the defects. However, the change in the energy is localized at the dislocations, and the reconstructed atomic density remains unaltered. The effect on the entire GBs was also addressed and directly reflects what was observed on the single dislocations. The values of the grain-boundary energy per unit length $F(\theta)/L$ increases with $\beta$, but the qualitative behavior is not significantly influenced, i.e. the same physical effects are accounted for, with different energies tuned by the additional term proposed here. 

The tuning of the energy of defects in a strained system was also discussed. In particular, the effect of the additional energy term on dislocations forming at the interface between layers with opposite in-plane strain was illustrated.  In agreement with previous cases, the energy increase was also found to be linear in $\beta$ for small $\beta$.  While the investigation of interfaces and tilted systems focused only on triangular or honeycomb structures, in this case fcc and bcc crystal symmetries were considered.  Indeed, this investigation was exploited to show the applicability of the general approach to 3D systems. The study of more crystal symmetries illustrates the generality of the APFC equations as discussed in
Sect.~\ref{sec:APFC} with all the complementary details provided in Appendix \ref{app:A}.

Overall, the proposed extension of the energy, Eq.~\eqref{eq:coreenergy}, allows one to control the energy of defects locally without changing their structure or general behavior. That is, elastic properties and defect energies may be tuned easily and independently. This becomes important when studying the competition of elastic and plastic relaxation in materials using APFC. 

The simulations reported in this work were performed using a FEM approach that deeply exploits mesh adaptivity. A semi-implicit time discretization scheme has been adopted, and it is reported in Appendix \ref{app:B1}. Dedicated work will be devoted to further optimize the method and provide more efficient calculations, useful for further extensions of the APFC model and to provide
extensive studies in 3D.

The modeling presented in this work is compatible with APFC approaches by considering the proper order parameters and the coupling with other effects. For instance, it would be interesting to examine the tuning of defects energies and interfaces in binary systems as in Refs.~\cite{ElderPRE2010,ElderJPCM2010} or at GBs when compositional domains are also present \cite{Geslin2015,Xu2016}.

\section*{Acknowledgements}
M.S. acknowledges the support of the Postdoctoral Research Fellowship awarded by the Alexander von Humboldt Foundation. R.B. and A.V. acknowledge the financial support from the German Research Foundation (DFG) under Grant No. SPP 1959. K.R.E. acknowledges financial support from the National Science Foundation under Grant No. DMR1506634 and the DRESDEN Fellowship Programme. The computational resources were provided by ZIH at TU Dresden and by the J\"ulich Supercomputing Center within Project No. HDR06.

\appendix
\section{Symmetry-dependent terms in the amplitude equations}
\label{app:A}

In Sect.~\ref{sec:APFC} the APFC model is presented. By exploiting the long-wavelength limit for the amplitudes \cite{Yeon2010,ElderPRE2010}, a
general structure for the amplitude equation can be derived independently of the crystal symmetry except for the definition of $f^{s}(\{\eta_j\},\{\eta_j^*\})$ (hereafter just $f^{s}$). For the sake of readability, the general approach is reported in the main text, while the details related to the specific lattice structure are reported in this appendix.

\subsection{Triangular or honeycomb 2D symmetry}
The reciprocal-space vectors are
\begin{equation}
\begin{gathered}
\mathbf{k}_1=k_0 \left(-\sqrt{3}/2,-1/2 \right),\ \mathbf{k}_2=k_0(0,1), \\
\mathbf{k}_3=k_0\left(\sqrt{3}/2,-1/2 \right), 
\end{gathered}
\label{eq:ktri}
\end{equation}
with $k_0=1$.
The term in the energy functional \eqref{eq:energyamplitude} reads
\begin{equation}
f^{\rm tri} = -2t \left(\eta_1\eta_2\eta_3 + c.c.\right),
\label{eq:fftri}
\end{equation}
while the corresponding contribution to the evolution laws for $\eta_j$ in Eq.~\eqref{eq:amptimefuncder} is
\begin{equation}
\frac{\delta f^{\rm tri}}{\delta \eta_j} = -2t \prod_{i\neq j}^3\eta_i^*.  
\label{eq:dfftri}
\end{equation}
The constant value of the amplitudes for an equilibrium crystal is,
\begin{equation}
\phi_0^{\rm tri} = \frac{t\pm\sqrt{t^2-15v\Delta B_0}}{15v},
\label{eq:phitri}
\end{equation}
as obtained from the minimization of Eq.~\eqref{eq:energyamplitude} with respect to $\eta_j=\phi_0$ with $f^s=f^{\rm tri}$.  The $+$ solution is valid for $t>0$, which produces a triangular array of maxima and the $-$ solution is valid for $t<0$ which produces a honeycomb array of maxima.

\subsection{Bcc symmetry}
The reciprocal-space vectors are
\begin{equation}
\begin{gathered}
\mathbf{k}_1=k_0 \left(1,1,0\right),\ \mathbf{k}_2=k_0\left(1,0,1\right), \\
\mathbf{k}_3=k_0\left(0,1,1\right),\ \mathbf{k}_4=k_0 \left(0,1,-1\right),\\
\mathbf{k}_5=k_0\left(1,-1,0\right),\ \mathbf{k}_6=k_0\left(-1,0,1\right).
\end{gathered}
\label{eq:kbcc}
\end{equation}
with $k_0=\sqrt{2}/2$. 
The term in the energy functional \eqref{eq:energyamplitude} reads
\begin{equation}
\begin{split}
f^{\rm bcc} =& -2t(\eta_1^*\eta_2\eta_4+\eta_2^*\eta_3\eta_5+\eta_3^*\eta_1\eta_6 + \eta_4^*\eta_5^*\eta_6^*+c.c.)\\
	     & +6v(\eta_1\eta_3^*\eta_4^*\eta_5^*+\eta_2\eta_1^*\eta_5^*\eta_6^*+\eta_3\eta_2^*\eta_6^*\eta_4^*+c.c.). 
\end{split}
\label{eq:ffbcc}
\end{equation}
The corresponding contributions in Eq.~\eqref{eq:amptimefuncder} can be written as
\begin{equation}
\begin{split}
\frac{\delta f^{\rm bcc}}{\delta \eta_i^*} =& -2t(\eta_k\eta_n^*+\eta_j\eta_l)+6v(\eta_k\eta_l\eta_m+\eta_j\eta_m^*\eta_n^*), \\
\frac{\delta f^{\rm bcc}}{\delta \eta_l^*} =& -2t(\eta_m^*\eta_n^*+\eta_i\eta_j^*)+6v(\eta_i\eta_k^*\eta_m^*+\eta_k\eta_j^*\eta_n^*), \\
\end{split}
\label{eq:ffbcc2}
\end{equation}
where all the equations for the amplitudes are obtained by permutations on the groups $(i,j,k)=(1,2,3)$ and $(l,m,n)=(4,5,6)$.

The constant value of the amplitudes in equilibrium is, 
\begin{equation}
\phi_0^{\rm bcc} = \frac{2t+\sqrt{4t^2-45v\Delta B_0}}{45v},
\label{eq:phibcc}
\end{equation}
as obtained from the minimization of Eq.~\eqref{eq:energyamplitude} with respect to $\eta_j=\phi_0$ with $f^s=f^{\rm bcc}$.

\subsection{Fcc symmetry}
The reciprocal-space vectors are
\begin{equation}
\begin{gathered}
\mathbf{k}_1=k_0 \left(-1,1,1\right),\ \mathbf{k}_2=k_0\left(1,-1,1\right), \\
\mathbf{k}_3=k_0\left(1,1,-1\right),\ \mathbf{k}_4=k_0 \left(-1,-1,-1\right),\\
\mathbf{k}_5=k_0\left(2,0,0\right),\ \mathbf{k}_6=k_0\left(0,2,0\right),\\
\mathbf{k}_7=k_0\left(0,0,2\right).
\end{gathered}
\label{eq:kfcc}
\end{equation}
with $k_0=\sqrt{3}/3$. Notice that at variance from triangular or bcc symmetry, two different sets of vectors with different length are present in Eq.~\eqref{eq:kfcc}.  This has to be taken into account when considering the $|\mathbf{k}_j|^2$ factor of Eq.~\eqref{eq:amplitudetime} which is equal to $4/3$ for $\mathbf{k}_{5,6,7}$ while it is 1 in all the other case (also with regard to other symmetries).  The term in the energy functional \eqref{eq:energyamplitude} reads
\begin{equation}
\begin{split}
f^{\rm fcc} =
&-2t[\eta_1^*(\eta_2^*\eta_5+\eta_3^*\eta_7+\eta_4^*\eta_6^*)+\eta_2^*(\eta_3^*\eta_6+\eta_4^*\eta_7^*)\\
&+\eta_3^*\eta_4^*\eta_5^*+c.c.] + 6v [\eta_1^*(\eta_2^*\eta_3^*\eta_4^*
+\eta_2 \eta_6^* \eta_7  + \eta_3\eta_5 \eta_6^*\\ 
	      &+ \eta_4\eta_5\eta_7) + \eta_2^*\eta_5(\eta_3\eta_7^*+\eta_4\eta_6) + \eta_3^*\eta_4\eta_6\eta_7+c.c.]. \\
\end{split}
\label{eq:fffcc}
\end{equation}
The contributions to Eq.~\eqref{eq:amptimefuncder} are
\begin{equation}
\begin{split}
\frac{\delta f^{\rm fcc}}{\delta \eta_1^*} =& 6v(\eta_2^*\eta_3^*\eta_4^* + \eta_2\eta_6^*\eta_7 + \eta_3\eta_5\eta_6^* + \eta_4\eta_5\eta_7) \\
					    & -2t(\eta_2^*\eta_5+\eta_3^*\eta_7+\eta_4^*\eta_6^*), \\
\frac{\delta f^{\rm fcc}}{\delta \eta_2^*} =& 6v(\eta_1^*\eta_3^*\eta_4^* + \eta_1\eta_6\eta_7^* + \eta_3\eta_5\eta_7^* + \eta_4\eta_5\eta_6) \\
					    & -2t(\eta_3^*\eta_6+\eta_4^*\eta_7^*+\eta_1^*\eta_5), \\				    
\frac{\delta f^{\rm fcc}}{\delta \eta_3^*} =& 6v(\eta_1^*\eta_2^*\eta_4^* + \eta_1\eta_5^*\eta_6 + \eta_2\eta_5^*\eta_7 + \eta_4\eta_6\eta_7) \\
					    & -2t(\eta_4^*\eta_5^*+\eta_1^*\eta_7+\eta_2^*\eta_6), \\
\frac{\delta f^{\rm fcc}}{\delta \eta_4^*} =& 6v(\eta_1^*\eta_2^*\eta_3^* + \eta_1\eta_5^*\eta_7^* + \eta_2\eta_5^*\eta_6^* + \eta_3\eta_6^*\eta_7^*) \\
					    & -2t(\eta_1^*\eta_6^*+\eta_2^*\eta_7^*+\eta_3^*\eta_5^*), \\
\frac{\delta f^{\rm fcc}}{\delta \eta_5^*} =& 6v(\eta_1\eta_3^*\eta_6 + \eta_2\eta_4^*\eta_6^* + \eta_2\eta_3^*\eta_7 + \eta_1\eta_4^*\eta_7^*) \\
					    & -2t(\eta_1\eta_2+\eta_3^*\eta_4^*), \\
\frac{\delta f^{\rm fcc}}{\delta \eta_6^*} =& 6v(\eta_1^*\eta_2\eta_7 + \eta_3\eta_4^*\eta_7^* + \eta_1^*\eta_3\eta_5 + \eta_2\eta_4^*\eta_5^*) \\
					    & -2t(\eta_2\eta_3+\eta_1^*\eta_4^*), \\
\frac{\delta f^{\rm fcc}}{\delta \eta_7^*} =& 6v(\eta_2^*\eta_3\eta_5 + \eta_1\eta_4^*\eta_5^* + \eta_1\eta_2^*\eta_6 + \eta_3\eta_4^*\eta_6^*) \\
					    & -2t(\eta_1\eta_3+\eta_2^*\eta_4^*). \\					    					   
\end{split}
\label{eq:fffcc2}
\end{equation}
Under the assumption of identical amplitudes, $\phi_0$ is,
\begin{equation}
\phi_0^{\rm fcc} = \frac{18t+\sqrt{324t^2-3087v\Delta B_0}}{441v},
\label{eq:phifcc}
\end{equation}
as obtained from the minimization of Eq.~\eqref{eq:energyamplitude} with respect to $\eta_j=\phi_0$ with $f^s=f^{\rm fcc}$. However, even when considering relaxed crystal with real and constant amplitudes, $\eta_j$ with $j \le 4$ and with $j\ge 5$ are not equivalent in Eq.~\eqref{eq:fffcc}. By assuming
\begin{equation}
\eta_j = \left\{ 
\begin{split}
\xi& \ \ j\le4 \\
\psi& \ \ j\ge5
\end{split}
\right.
\end{equation}
we can write the stationary conditions $\delta F/\delta \xi=0$ and $\delta F/\delta \psi=0$ and solve for $\xi$ and $\psi$.  For the parameter adopted in Sect.~\ref{sec:strained}, we calculated $\xi=1.334$ and $\psi=1.002$, which are used to set the initial conditions for strained fcc crystals by means of $\phi_0=\phi_{0,j}$ in Eq.~\eqref{eq:etaelas}.

\section{ FEM implementation}
\label{app:B}
\subsection{Discretization scheme}
\label{app:B1}
The calculation of the evolution in time of $\eta_j$ has been performed by considering different equations for their real and imaginary parts. The following array of functions is considered $\boldsymbol{\alpha}=[\text{Re}(\eta_1),\text{Im}(\eta_1),...,\text{Re}(\eta_{N}),\text{Im}(\eta_{N})]$, indexed by $p=1,...,2N$ and we define $k=2j-1$. With this choice $\eta_j = \alpha_k+i \alpha_{k+1}$. Moreover, we split the fourth-order PDE in \eqref{eq:amplitudetime} in two second-order PDEs, namely
for $\partial \eta_j/ \partial t$ and $\mathcal{G}_j\eta_j=\zeta_k+i\zeta_{k+1}$. The resulting four equations read 
\begin{equation}
\begin{split}
\frac{\partial \alpha_k}{\partial t}=&-|\mathbf{k}_j|^2 \bigg[ \Delta B_0
\alpha_k +B_0^x\nabla^2 \zeta_{k} -2B_0^x\mathbf{k}_j \cdot \nabla \zeta_{k+1}
\\ &+ 3v (A^2-|\eta_j|^2) \alpha_k + \text{Re}\left(\frac{\delta f^{s}}{\delta \eta_j^*}\right) \bigg], \\
\frac{\partial \alpha_{k+1}}{\partial t}=&-|\mathbf{k}_j|^2 \bigg[\Delta B_0 \alpha_{k+1} +B_0^x\nabla^2 \zeta_{k+1} +2B_0^x\mathbf{k}_j \cdot \nabla \zeta_{k} \\ 
&+ 3v(A^2-|\eta_j|^2) \alpha_{k+1} + \text{Im}\left(\frac{\delta f^{s}}{\delta \eta_j^*}\right) \bigg], \\
\zeta_k =& \nabla^2\alpha_k - 2\mathbf{k}_j \cdot \nabla \alpha_{k+1}, \\
\zeta_{k+1} =& \nabla^2\alpha_{k+1} + 2\mathbf{k}_j\cdot \nabla \alpha_{k}.
\end{split}
\label{eq:eqeq}
\end{equation}

Let us consider the time discretization $t_n$ with $n \in \mathbb{N}$ such as $0=t_0<t_1< ...$ and the timestep $\tau_n=t_{n+1}-t_{n}$. The adopted semi-implicit integration scheme in the matrix form reads $\mathbf{L} \cdot \mathbf{x}=\mathbf{R}$ with

\begin{equation}
\mathbf{L}=
\begin{bmatrix}
-\nabla^2 & \mathcal{A} &1 &0 \\[0.75em]
-\mathcal{A} &-\nabla^2 &0 &1 \\[0.75em]
G_1(\{\alpha_i^{(n)}\}) & 0 & \mathcal{K} \nabla^2 & -\mathcal{K} \mathcal{A} \\[0.75em]
0 & G_2(\{\alpha_i^{(n)}\}) & \mathcal{K} \mathcal{A} & \mathcal{K} \nabla^2  
\end{bmatrix}
\label{eq:integrationschemeL}
\end{equation}
\begin{equation}
 \mathbf{x}=
\begin{bmatrix}
\alpha_k^{(n+1)} \\[0.75em]
\alpha_{k+1}^{(n+1)} \\[0.75em] 
\zeta_k^{(n+1)}  \\[0.75em]
\zeta_{k+1}^{(n+1)}
\end{bmatrix} \qquad
\mathbf{R}= 
\begin{bmatrix}
0\\[0.75em]
0\\[0.75em] 
H_1(\{\alpha_i^{(n)}\})\\[0.75em]
H_2(\{\alpha_i^{(n)}\}) \\
\end{bmatrix}
\label{eq:integrationschemexR}
\end{equation}
where $\mathcal{A}=2\mathbf{k}_j \cdot \nabla$ and $\mathcal{K}=|\mathbf{k}_j|^2B_0^x$, while the functions evaluated explicitly at time $t_n$ are given by
\begin{equation}
\begin{split}
G_1(\{\alpha_i\})=&\dfrac{1}{\tau_n}+|\mathbf{k}_j|^2\Delta B+3v|\mathbf{k}_j|^2\left(A^2+\alpha_k^2-\alpha_{k+1}^2\right),\\
G_2(\{\alpha_i\})=&\dfrac{1}{\tau_n}+|\mathbf{k}_j|^2\Delta B+3v|\mathbf{k}_j|^2\left(A^2+\alpha_{k+1}^2-\alpha_{k}^2\right), \\
H_1(\{\alpha_i\})=&\left[ \dfrac{1}{\tau_n}+6|\mathbf{k}_j|^2v \alpha_k^2 \right]\alpha_k -|\mathbf{k}_j|^2\text{Re}\left(\frac{\delta f^{s}}{\delta \eta_j^*}\right), \\
H_2(\{\alpha_i\})=&\left[\dfrac{1}{\tau_n} +6|\mathbf{k}_j|^2v \alpha_{k+1}^2 \right]\alpha_{k+1} -|\mathbf{k}_j|^2\text{Im}\left(\frac{\delta f^{s}}{\delta \eta_j^*}\right).
\end{split}
\label{eq:system}
\end{equation}

The functions in \eqref{eq:system} account for the right- and left-hand side terms resulting from the linearization of $-3v \left( |A|^2-|\eta_j|^2 \right)\alpha_k$ and $-3v \left( |A|^2-|\eta_j|^2 \right)\alpha_{k+1}$ terms in \eqref{eq:eqeq} as function of $\alpha_k^{(n+1)}$ and $\alpha_{k+1}^{(n+1)}$ around $\alpha_k^{(n)}$ and $\alpha_{k+1}^{(n)}$, respectively \cite{Raetz2006}. The ordering of the equations in the system is adopted in order to have the $\nabla^2$ term along the diagonal. This allows for high efficiency in the calculation of the numerical solution, in particular when using iterative solvers. In order to compute the evolution of the amplitudes from Eq.~\eqref{eq:amplitudetime} the system defined by \eqref{eq:integrationschemeL} and \eqref{eq:integrationschemexR} has to be solved for each $\eta_j$, i.e. a number of (coupled) systems equal to the number of different amplitude functions (i.e. $\mathbf{k}_j$ vectors) has to be
considered.

So far, only the implementation of the standard APFC model has been considered. The contribution introduced in Sect.~\ref{sec:coreenergy}, providing the additional term in the evolution laws as reported in Eq.~\eqref{eq:evolcoreenergy}, is readily included by computing the quantity $\nabla^2 A^2$ and adding the term $\beta \nabla^2 A^2$ to the matrix \eqref{eq:integrationschemeL} at $\mathbf{L}_{31}$ and $\mathbf{L}_{42}$.

The integration scheme reported in this appendix has been implemented in the Finite Element Method framework available within the AMDiS toolbox \cite{Vey2007,Witkowski2015}.

\subsection{Spatial Adaptivity}
\label{app:B2}
As mentioned in Sec.~\ref{sec:num}, an adaptive spatial discretization has been adopted in order to optimize the numerical simulations. In particular, we considered a refinement of the computational grid where the real and complex parts of $\eta_j$ oscillate. Notice that according to the specific ${\mathbf{k}_j}$ vectors and the deformation of the crystal, amplitude functions may oscillate differently. Here, we detect the region where oscillations occur by evaluating where the quantity $\sum_{j=1}^N |\nabla [\text{Im}(\eta_j)]|$ is non-vanishing (over an arbitrary threshold), and we set the refinement to ensure proper resolution for all the $\eta_j$ functions. In addition to this criterion, the spatial discretization is further refined where non-vanishing values of $|{\nabla}A^2|$ are present in order to ensure the proper resolution also at defects and interfaces, typically involving changes in the amplitudes on smaller lengthscales than in the bulk. At variance with the work reported in Ref.~\cite{AthreyaPRE2007} a change in the equations of the APFC model is not required here. 

\subsection{Simulation Domain for Periodic Boundaries}
\label{app:B3}
In order to simulate infinitely extended, tilted crystal with periodic boundary conditions, matching amplitudes have to be set at the boundaries of the simulation domain. For the setup adopted in Sec.~\ref{sec:tilted} this occurs at the boundaries perpendicular to the GBs.

For the GBs shown in Fig.~\ref{fig:figure4}(a)-(c), the size of the domain along $\hat{\mathbf{x}}$ is set to $L_x = 320\pi$. To avoid
the presence of a further discontinuity in the crystal orientation at the top and bottom boundaries, $L_y$ has to be set according to the specific choice of
$\theta$, ensuring matching amplitudes at the boundaries with normal along $\hat{\mathbf{y}}$.  
Therefore, by considering the triangular symmetry, $\mathbf{k}_j$ vectors reported in \eqref{eq:ktri} and the tilt angle $\theta$ affecting the amplitudes as from Eq.~\eqref{eq:krot}, $L_y$ must be an integer number of
\begin{equation}
\begin{split}
\lambda_1=&\frac{4\pi}{-\sqrt{3}\sin\theta+1-\cos\theta}, \\
 \lambda_2=&\frac{2\pi}{1-\cos\theta}, \\
 \lambda_3=&\frac{4\pi}{\sqrt{3}\sin\theta+1-\cos\theta}. 
\end{split}
\label{eq:size1}
\end{equation}
$\lambda_2$ is the largest wavelength along $\hat{\mathbf{y}}$ for small $\theta$. In the simulations reported in the following we select some $\theta$ values for which $\lambda_2(\theta)/\lambda_1(\theta)$ and $\lambda_2(\theta)/\lambda_3(\theta)$ gives an integer number. Then, $L_y=\lambda_2(\theta)$. 

To simulate an horizontal GB, as in Fig.~\ref{fig:figure4}(e)-(g), the size of the domain along $\hat{\mathbf{y}}$ is set to $L_y = 320\pi$. Then, matching amplitudes have to be set at the boundaries with normal along $\hat{\mathbf{x}}$. $L_x$ must then be an integer number of \begin{equation}
\begin{split}
\lambda_1'=&\frac{4\pi}{\sin\bar{\theta}-\sqrt{3}\cos(\bar{\theta}-1)}, \\ 
\lambda_2'=&\frac{2\pi}{\sin\bar{\theta}}, \\
\lambda_3'=&\frac{4\pi}{\sin\bar{\theta}+\sqrt{3}(\cos\bar{\theta}-1)}. 
\end{split}
\label{eq:size2}
\end{equation}
$\lambda_3'$ is the largest wavelength in the $\hat{\mathbf{x}}$ direction for small $\bar{\theta}$. However, at variance from the vertical GB it has a value comparable to the others for small $\bar{\theta}$. Therefore, $\bar{\theta}$ values are chosen in order to have integer numbers for $M\lambda_3'(\bar{\theta})/\lambda_1'(\bar{\theta})$
and $M\lambda_3'(\bar{\theta})/\lambda_2'(\bar{\theta})$, with $M$ significantly larger than 1. Then, $L_x=M\lambda_3'(\bar{\theta})$. 
\\

\end{document}